\documentclass[11pt]{article}
\usepackage{cite}
\usepackage{amsmath,amsfonts,amssymb}
\usepackage[small,bf,hang]{caption}
\usepackage{slashed}

\makeatletter\let\corresponds\@undefined\makeatother
\usepackage{mathabx}

\usepackage[vcentermath]{youngtab}

\def\hybrid{
        \topmargin -20pt
        \oddsidemargin 0pt
        \headheight 0pt \headsep 0pt
        \textwidth 6.25in 
        \textheight 9.5in 
        \marginparwidth .875in
        \parskip 5pt plus 1pt \jot = 1.5ex}

\hybrid

 \csname
@addtoreset\endcsname{equation}{section}

\newcommand{\vev}[1]{\langle #1 \rangle}
\newcommand{\nin}[1] {\underline{\phantom{h}}\hskip-6pt {#1}}

\def\moth{\mathsurround=0pt}
\newdimen\zo \zo=0pt

\def\tick{\leaders\hrule height 0.5ex depth 0pt \hskip 0.5pt}
\def\upboxfill{$\moth \setbox\zo\hbox{\tick}%
  \hskip 3pt\hbox to 0pt{$\tick$\hss}\hrulefill \hbox to 7.5pt{$\tick$\hss}$}

\def\dtick{\leaders\hrule height .34pt depth 0.5ex \hskip 0.5pt}
\def\downboxfill{$\moth \setbox\zo\hbox{\dtick}%
  \hskip 2pt\hbox to 0pt{$\dtick$\hss}\hrulefill \hbox to 2pt{$\dtick$\hss}$}

\def\bec{\begin{center}}
\def\ec{\end{center}}

 \def\det{{\rm det\,}}
\def\be{\begin{equation}}
\def\ee{\end{equation}}
\def\bea{\begin{eqnarray}}
\def\eea{\end{eqnarray}}
\def\ba{\begin{array}}
\def\ea{\end{array}}

\begin{document}

\begin{titlepage}
\rightline{}
\rightline{\tt MIT-CTP-4428}
\rightline{\tt  LMU-ASC 80/12}
\rightline{December 2012}
\begin{center}
\vskip 2.5cm
{\Large \bf {Towards an invariant geometry of double field theory}}\\
\vskip 2.5cm
{\large {Olaf Hohm${}^1$ and Barton Zwiebach${}^2$}}
\vskip 1cm
{\it {${}^1$Arnold Sommerfeld Center for Theoretical Physics}}\\
{\it {Theresienstrasse 37}}\\
{\it {D-80333 Munich, Germany}}\\
olaf.hohm@physik.uni-muenchen.de
\vskip 0.7cm
{\it {${}^2$Center for Theoretical Physics}}\\
{\it {Massachusetts Institute of Technology}}\\
{\it {Cambridge, MA 02139, USA}}\\
zwiebach@mit.edu

\vskip 1.5cm
{\bf Abstract}
\end{center}

\vskip 0.4cm

\noindent
\begin{narrower}
We introduce a 
geometrical framework for double field theory 
in which generalized Riemann and torsion tensors are defined without reference to a
particular basis. This invariant geometry provides a unifying framework for the 
frame-like and metric-like formulations developed before. We discuss the relation to 
generalized geometry and  give an 
`index-free' proof of the algebraic Bianchi identity. 
Finally, we analyze to what extent the 
generalized Riemann tensor encodes the curvatures of Riemannian 
geometry. We show that it contains the  conventional 
Ricci tensor and scalar curvature but not 
the full Riemann tensor, suggesting the possibility of a further extension of this
framework. 

\end{narrower}

\end{titlepage}

\newpage

\tableofcontents

\newpage

\section{Introduction}
 Double field theory is a framework to render the 
T-duality group $O(D,D)$ a manifest symmetry for the low-energy effective spacetime actions of string theory.  
This is achieved  
by introducing doubled coordinates $X^{M}=(\tilde{x}_i,x^i)$, $M,N=1,\ldots, 2D$,
namely 
by augmenting the usual spacetime 
coordinates $x^i$, $i=1,\ldots, D$, by an equal number 
of new `winding-type' coordinates $\tilde{x}_i$ \cite{Hull:2009mi,Hull:2009zb,Hohm:2010jy,
Hohm:2010pp}.  
The massless fields of bosonic string theory, the metric $g_{ij}$, the 2-form 
$b_{ij}$ and the scalar dilaton $\phi$, are encoded by novel geometrical objects 
that are tensors under $O(D,D)$. A generalized metric ${\cal H}_{MN}$ that is a 
symmetric $O(D,D)$ matrix encodes $g_{ij}$ and $b_{ij}$, and an $O(D,D)$ singlet 
$d$ encodes the scalar dilaton $\phi$  
 via $e^{-2d}=\sqrt{-g}e^{-2\phi}$.   
An $O(D,D)$ and gauge invariant spacetime action for double field theory can then be 
written without any reference to the original fields $g$, $b$ and $\phi$.  
This theory has been originally formulated in \cite{Hull:2009mi,Hull:2009zb,Hohm:2010jy,
Hohm:2010pp}. Earlier important work can be found in \cite{Siegel:1993th,Tseytlin:1990nb,Duff:1989tf}
and further developments have been discussed in \cite{Hohm:2010xe,Kwak:2010ew,Hohm:2011gs,Hohm:2011dz,Hohm:2011ex,Hohm:2011zr,Hohm:2011cp,Hohm:2011nu,Hohm:2011si,Hillmann:2009ci,Berman:2010is,Berman:2012uy,Berman:2012vc, 
Jeon:2010rw,Jeon:2011cn,Jeon:2011sq,Schulz:2011ye,Copland:2011yh,
Thompson:2011uw,Albertsson:2011ux,Andriot:2011uh,grana-marques,Andriot:2012wx}.

In this paper we aim to take the first steps towards a fully invariant formulation of the 
geometry of double field theory, by which we mean a formulation that does not require 
the introduction of a coordinate basis.  
There are two aspects to this problem. First, the notion of manifold 
needs to be generalized   
because the gauge transformations are not given 
by diffeomorphisms of the doubled space. Second, 
we need to introduce invariant curvatures 
that are compatible with these novel gauge symmetries and that allow us to define an invariant action. 

In order to explain the first part of the problem we recall  
the infinitesimal gauge transformations of double field theory, which are parameterized by 
an $O(D,D)$ vector parameter $\xi^{M}$ and read 
 \be\label{delH}
 \begin{split}
   \delta_{\xi}{\cal H}_{MN} \ &= \ 
  \xi^{P}\partial_{P}{\cal H}_{MN}+\big(\partial_{M}\xi^{P}-\partial^{P}\xi_{M}\big)
  {\cal H}_{PN}+\big(\partial_{N}\xi^{P}-\partial^{P}\xi_{N})
  {\cal H}_{MP}\;,   \\
  \delta_{\xi}\big(e^{-2d}\big) \ &= \ \partial_M\big(\xi^{M}e^{-2d}\big)\;,  
 \end{split}
 \ee
where indices are raised and lowered by the $O(D,D)$ invariant metric 
 \be\label{eta}
  \eta_{MN} \ = \ \begin{pmatrix}    0 & {\bf 1} \\[0.5ex]
  {\bf 1} & 0 \end{pmatrix}\;.
   \ee
We infer from (\ref{delH}) that $e^{-2d}$ transforms as a scalar density and so can be treated as 
in ordinary differential geometry and be used to define an invariant integration. In contrast, 
the gauge transformation of the generalized metric 
does not take the form of  a 
Lie derivative in the doubled space 
but rather defines a generalized Lie derivative $\widehat{\cal L}_{\xi}$
by the relation 
$\widehat{\cal L}_{\xi}{\cal H}_{MN} \equiv \delta_{\xi}{\cal H}_{MN}$.  
These generalized Lie derivatives leave the $O(D,D)$ metric invariant, 
$\widehat{\cal L}_{\xi}\eta_{MN}=0$ \cite{Hohm:2010pp}. 
Since 
we cannot think of (\ref{delH}) as 
an infinitesimal general coordinate transformation, 
 we have to define suitably generalized 
coordinate transformations. 
A generalized notion of manifold is required in which   
the transition between different coordinate patches is governed by these generalized 
coordinate transformations 
and so that there is a well-defined \textit{constant} metric (\ref{eta}).  
This part of the problem has recently been addressed 
 by us in \cite{Hohm:2012gk}
and will be briefly reviewed in sec.~\ref{secfinite}. In this paper we will be mainly concerned with 
the second part of the problem, and thus the present paper can be seen as a companion to \cite{Hohm:2012gk}.

The second part of the problem requires the introduction of invariant curvatures on 
the generalized (doubled) manifold.  This should be possible, because the action of double field theory can be 
written in a geometric Einstein-Hilbert-like 
form,  
  \be\label{DFTaction}
  S_{\rm DFT} \ = \  \int dxd\tilde{x}\,e^{-2d}\,{\cal R}({\cal H},d)\;, 
 \ee
where ${\cal R}$ is an $O(D,D)$ scalar and a gauge scalar and can thus be viewed as a generalized 
curvature scalar. Similarly, the variation of (\ref{DFTaction}) with respect to ${\cal H}_{MN}$ gives 
a tensor ${\cal R}_{MN}$ that transforms covariantly under gauge transformations, i.e., with the 
generalized Lie derivative as in (\ref{delH}), and   
can be viewed as a generalized Ricci tensor. 
It is then natural to seek an analogue to Riemannian geometry, so that by introducing connections and 
invariant curvatures  one can systematically construct the Ricci tensor and curvature scalar. 
There indeed exist two such formulations which have been developed in physicist terminology, i.e., 
defining  everything with respect to a basis and introducing `index-based' objects.  
The first formalism was 
developed  
some time ago by Siegel \cite{Siegel:1993th}.  
It is a frame formalism that is the analogue of the vielbein formulation of general relativity and 
 has been related to double field theory in~\cite{Hohm:2010xe}. 
The second formulation is a metric-like formalism with Christoffel-type  connections. 
As in general relativity, the second formulation is related to the first by a `vielbein postulate' \cite{Hohm:2010xe,Hohm:2011si}. 
It has  been developed in a self-contained fashion in \cite{Hohm:2011si}, 
using elements of one of the formulations 
of Jeon, Lee, and Park \cite{Jeon:2010rw,Jeon:2011cn}, to which
it reduces upon performing a (non-covariant) truncation. 

Our aim in this paper is to provide `invariant' definitions of the generalized Riemann and torsion tensors.
While these definitions require the use of a basis of vector fields, this basis is totally arbitrary. The 
tensors are manifestly 
independent of this choice and thus basis independent. Our formulation does not require the 
use of a coordinate basis nor of a frame basis with further constraints. Specifically,   
we wanted to find the
analogue of the well-known definition of the Riemann tensor in ordinary geometry:
\be
\label{theoriginal}
{R} (x, y, z, w)  \ \equiv  \  \vev{(\nabla_x \nabla_y -  \nabla_y \nabla_x- \nabla_{[x, y]} ) z \,, w}\,.
\ee
In here $x, y, z, w$ are sections on the tangent bundle of the manifold (vector fields), $\nabla$ 
is a connection and $\langle \cdot \,, \cdot \rangle$ is an inner product on the tangent bundle.\footnote{In Riemannian geometry a more basic definition of the Riemann tensor  
does not use the metric.  The Riemann tensor is viewed as   
a linear operator $R(x,y)$ defined to act on vector fields as $R(x,y)z =   (\nabla_x \nabla_y -  \nabla_y \nabla_x- \nabla_{[x, y]} ) z$.  There seems to be no 
analogue of this metric-independent definition in a doubled geometry.} 
We found such formula.  With $X, Y, Z, W$ denoting generalized vector fields, or sections of 
a suitably generalized `tangent bundle' 
to the doubled manifold,  $\nabla$ a connection, 
and inner product $\langle X, Y\rangle = \eta_{MN} X^M Y^N$,  
we define the generalized Riemann tensor  by
\be
\label{curvaturehereIntro}  
\begin{split}
\phantom{\Bigl(}~~{\cal R} (X, Y, Z, W)  \ \equiv & \ \ \  \vev{(\nabla_X \nabla_Y -  \nabla_Y \nabla_X- \nabla_{[X, Y]_D} ) Z \,, W} \\
& \ \  + \vev{(\nabla_Z \nabla_W -  \nabla_W \nabla_Z - \nabla_{[Z, W]_D}) X \,, Y}  ~~\\
& \ \  +  \sum_{A}\,\vev{Y, \nabla_{Z_A} X}  \, \vev{W, \nabla_{Z^A} Z }\;. \phantom{\Bigl(}
\end{split}
\ee
Here $Z_{A}$ denotes an arbitrary basis of vector fields, 
with duals $Z^A$ such that
$\langle Z_A, Z^B\rangle = \delta_A{}^{B}$.  
In addition,  we use the so-called D-bracket 
that generates generalized Lie derivatives and is 
the double field theory extension of the Dorfman bracket. 
The first line formally coincides with the definition of the 
conventional Riemann tensor in (\ref{theoriginal}), but with the 
Lie bracket replaced by the D-bracket.
This replacement is important 
since the Lie bracket of two generalized vectors is not a generalized vector, while the D-bracket
of two generalized vectors is a generalized vector.  Still, the first line alone fails to define a tensor.
The other two lines are needed.  
Similarly, we wanted to generalize the torsion tensor\footnote{In Riemannian geometry
a more basic 
 definition does not use the metric and defines the vector field  
$T(x,y)=\nabla_x y-\nabla_y x-[x,y]$.} 
 \be
  T(x,y,z) \ = \ \vev{ \nabla_{x}y-\nabla_{y}x-[x,y],z}\;.
 \ee 
We found a generalized torsion tensor which reads 
\be
\label{defgentorIntro}
 {\cal T} (X, Y, Z) \ 
\equiv  \    \vev{ \nabla_X Y - \nabla_Y X- [X, Y]_D\, , \,Z } + \vev{Y, \nabla_Z X}  \,. 
\ee
Again, the first term formally coincides with the conventional torsion tensor, with the 
Lie bracket replaced by the D-bracket, but the last term is needed to preserve the tensor character.

We may then specialize these definitions to either 
a coordinate or frame basis, and we will see that the generalized Riemann and torsion tensors reduce to those 
previously introduced in the metric-  and  frame-like formalisms. As such, this formulation provides 
a unifying framework that makes manifest the equivalence of the 
`index-based' approaches of \cite{Siegel:1993th,Hohm:2010xe} and \cite{Hohm:2011si}.  
We illustrate the strength of this formulation by giving a basis independent proof of the 
algebraic Bianchi identity for the generalized Riemann tensor.

We will also comment on the relation to results in the generalized geometry developed by 
Hitchin, Gualtieri and others \cite{Hitchin:2004ut,Gualtieri:2003dx,Gualtieri:2007bq,Coimbra:2011nw}.  
In fact, the generalized torsion (\ref{defgentorIntro}) is closely related to the torsion defined 
by Gualtieri~\cite{Gualtieri:2007bq}.   
To the best of our knowledge, however, the generalized Riemann tensor (\ref{curvaturehereIntro}) has not 
appeared in the mathematical literature. 

We use the opportunity to analyze the generalized Riemann tensor 
in somewhat more detail than in \cite{Hohm:2011si}. In particular, we discuss a way to derive new differential 
Bianchi identities, in the course of which we present some technically interesting new results. 
For instance, just like in ordinary geometry, the gauge transformations of the connection
can be written covariantly.  Indeed, we show that the infinitesimal gauge transformations of the connection components 
$\Gamma_{MNK}$
can be 
written in terms of the generalized Riemann tensor,  
 \be\label{covGAMMAIntro}
\delta_\xi \Gamma_{MNK}  \ = \ 
2\big( \nabla_M \nabla_{[N} \xi_{K]}   - \nabla_{[N} \nabla_{K]} \xi_M\big)   +\xi^{P}{\cal R}_{PMNK}\;,  
\ee
all  
written with respect to a coordinate basis. Similarly, for a frame basis $E_{A}$, 
the gauge transformations of the 
spin connection components $\omega$ are written as  
 \be\label{finaldelomegaIntro}
  \delta \omega_{MAB} \ = \ \big[\nabla_A,\nabla_B\big]\xi_M+{\cal R}_{ABMN}\xi^{N}\;. 
 \ee
Note that in the generalized geometric framework the right-hand side is non-zero, in contrast to 
conventional Riemannian geometry, where the commutator of covariant derivatives can be 
expressed in terms of the Riemann tensor. Even though there is no simple relation between the commutator of 
covariant derivatives and the Riemann tensor, we find an intriguing relation for a certain 
\textit{triple commutator} of covariant derivatives in terms of the generalized Riemann tensor and its 
covariant derivatives, see eq.~(\ref{triplecomm}) below.

We finally discuss the extent to which the generalized Riemann tensor encodes  the usual curvatures of Riemannian geometry. 
We confirm that it 
contains the Ricci tensor and Ricci scalar, but  
we also establish that 
it does not contain the full uncontracted Riemann tensor. 
This implies that while  
the present framework is satisfactory and sufficient for the two-derivative part of the 
effective action, 
the inclusion of higher-derivative $\alpha'$ corrections requires an extension of 
this geometry. 
We will argue in the conclusions that there are strong reasons to believe that  
$\alpha'$ corrections are possible in double field theory so that such an extension should exist. 

We believe that our results are a first step 
towards a properly invariant geometric  
framework underlying double field theory. Needless to say,   
there are various gaps to be filled in order to achieve a mathematically satisfactory 
formulation.  
One important aspect of double field theory that should be properly accounted for
 is related to the need to impose the following constraint  
 \be\label{constraint}
  \partial^{M}\partial_{M} \ \equiv \ \eta^{MN}\partial_M\partial_N \ = \ 0\;, 
 \ee
with $\eta_{MN}$ defined in (\ref{eta}), and 
acting on arbitrary fields and gauge parameters.   
In this form, sometimes referred to as the \textit{weak constraint}, it is a direct consequence of the 
level-matching constraint in closed string theory. In the double field theories constructed so far, however, 
a stronger form is required. 
Since the product of two functions satisfying (\ref{constraint}) does not
necessarily satisfy (\ref{constraint}), 
we demand  
that $\partial^M\partial_M$ also annihilates 
all products of fields, thus requiring   
\be  
\label{sconstraint}
\partial^M A\, \partial_M B \ = \ 0\, \qquad \forall  A,B\,,  
\ee
in order to have a closed algebra of functions. 
Thus we are restricting to a subalgebra of functions on the doubled space. Almost certainly some version 
of (\ref{constraint}) and (\ref{sconstraint}) 
must be part of any rigorous definition of a generalized manifold, and understanding 
this properly may give insight  
 into the geometric meaning of the 
 level-matching constraint in string theory.\footnote{See \cite{Hohm:2011cp,grana-marques} for situations that require only relaxed versions
 of these constraints.} 
 One consequence of this constraint is that we cannot think 
 of a (generalized) vector field $V^M$ as a differential operator $V=V^M\partial_M$ acting on this subalgebra,  
 since this operator 
is unchanged under $V^M\rightarrow V^M+\lambda\, \partial^M\chi$, with $\lambda$ and $\chi$ arbitrary, while such a change does affect the generalized vector.  
Thus, we leave for further work 
a proper invariant treatment of the nature of the `generalized tangent bundle,' 
and  
we hope that our results
motivate mathematicians to further develop this geometrical framework. 
A first proposal on the underlying geometrical formulation of double field theory has already appeared in 
the mathematical literature \cite{Vaisman:2012ke}, but it is clear that we are still lacking a complete picture.

\section{Generalities of double field theory}
We start by introducing some basic notions of double field theory, particularly 
the C and D brackets, which are the double field theory counterparts of the  
Courant and Dorfman brackets of generalized geometry and play a key role
in the gauge transformations. This serves as a brief review
and also sets the notation. Then we recall the invariant definition of tensors 
and set the stage for our later introduction of a torsion and Riemann tensor by showing, using our
recent results in \cite{Hohm:2012gk}, that tensors defined by means of the 
C and D brackets 
are indeed generalized tensors under finite transformations in the sense of \cite{Hohm:2012gk}.

\subsection{Generalized Lie derivatives, Courant and Dorfman brackets}\label{invfinite}
A basic object in double field theory is the $O(D,D)$ invariant metric $\eta$ defined in (\ref{eta}). 
For later use we 
introduce an invariant notation for this metric  
by writing 
\be\label{scalarpr}
\vev{X , Y} \ = \  \eta_{MN} X^M Y^N\;,  
\ee
where here and in the following $X$, $Y$, $Z$, etc., denote vector fields on the doubled space. 
In particular, we view the partial derivatives $\partial_M$ as a coordinate basis of vector fields 
and write
 \be
  \vev{\partial_M , \partial_N} \ = \  \eta_{MN}\;. 
 \ee 
In general we have a natural action of vector fields on functions,  
$f\rightarrow X(f)$, giving a new function:
\be
X(f) \equiv  X^M\partial_M f \,.
\ee
We stress, however, that in the context of double field theory 
a vector field is not uniquely determined 
by its action on functions because these  
satisfy the strong constraint 
(\ref{constraint}) and (\ref{sconstraint}). 
Thus, we cannot introduce 
vector fields as in ordinary differential geometry, and currently we do not know how to define
generalized vectors in an invariant or geometric fashion. Below we will define generalized vectors 
by their (generalized) coordinate transformations, leaving their proper invariant treatment 
for future work, but we stress that once generalized vectors are given, higher tensors can be defined 
completely invariantly, as we will discuss below.

Let us now turn to the generalized Lie derivatives that govern the gauge transformations of double field theory as in (\ref{delH})
and are compatible with the metric (\ref{scalarpr}).  
The generalized Lie derivative is defined on an 
$O(D,D)$ tensor $V^{M}{}_{N}$ as 
 \be\label{genLieder}
  \widehat{\cal L}_{\xi} V^{M}{}_{N} \ = \ \xi^{K}\partial_K V^{M}{}_{N}+\left(\partial^M\xi_K-\partial_K\xi^M\right)V^K{}_{N}
  +\left(\partial_N\xi^K-\partial^K\xi_N\right)V^M{}_{K}\;,
 \ee 
and similarly for tensors in arbitrary representations of $O(D,D)$. Here the $O(D,D)$ indices are raised and 
lowered with the metric $\eta$. It is thus easy to see that $\eta$ is indeed invariant under generalized 
Lie derivatives, $\widehat{\cal L}_{\xi}\eta=0$. We refer to $O(D,D)$ tensors 
transforming with the generalized Lie derivative under gauge transformations as generalized 
tensors. Note that the scalar product (\ref{scalarpr}) of two generalized vectors is then a 
generalized scalar. Moreover, the partial derivative of a scalar is a generalized vector  \cite{Hohm:2010xe}. 
 
The generalized Lie derivatives form an algebra that in turn defines the C-bracket,
which is an $O(D,D)$ invariant extension of the Courant bracket in generalized geometry. We have\footnote{We note that here 
we view $\widehat{\cal L}_{X}$ as an abstract operator. Viewing it as a field variation, thus acting on 
fields first, leads to a different sign on the right-hand side of this relation.} 
 \be
  \big[\widehat{\cal L}_{X},\widehat{\cal L}_{Y}\big] \ = \ \widehat{\cal L}_{[X,Y]_C}\;, 
 \ee 
where the C-bracket reads 
\be
\label{cbrakc}
[X, Y]_C  \ = \   [X, Y]  -{1\over 2}  X_M \vec{\partial} Y^M  
+  {1\over 2}  Y_M \vec{\partial} X^M\;. 
\ee 
Here $[\, ,\,]$ denotes the usual Lie bracket of vector fields, 
 \be\label{usualLie}
  [X,Y]^M \ = \ X^N\partial_N Y^M-Y^N\partial_N X^M\;,  
 \ee
and $\vec{\partial}$ is a short-hand notation for the partial derivative 
with an index raised by the metric. 
Thus, in components, 
the C-bracket reads 
\be
\label{cbrakcc}
[X, Y]_C^K  \ = \   [X, Y]^K  -{1\over 2}  X_M \partial^K Y^M  
+  {1\over 2}  Y_M \partial^K  X^M\;. 
\ee 
Because of the strong constraint the C-bracket acting on functions gives the same as the Lie
bracket 
\be
\label{confunction}
[X, Y]_C  f  \ = \   [X, Y] \,f  \,.
\ee
The C-bracket of two generalized vectors is also a generalized vector \cite{Hohm:2010xe}.

Another useful bracket, the D-bracket,  can be defined directly through 
the generalized Lie derivative and turns out to 
be an $O(D,D)$ covariant extension of the Dorfman bracket in generalized geometry. 
We define 
 \be\label{Ddef}
  \big[ X,Y\big]_{D} \ \equiv \ \widehat{\cal L}_{X}Y\;.
 \ee
Although this is not antisymmetric we continue referring to it as a bracket.   
It differs from the C-bracket by a generalized vector, so it is also a generalized vector:  
\be\label{firstDbracket}
[X, Y]_D \ = \ [X, Y]_C  + {1\over 2}  \vec{\partial}\,  \vev{X, Y} \;. 
\ee
Its component expression is therefore
\be
\label{dorfbrakcc}
[X, Y]_D^K  \ = \   [X, Y]^K    +   Y_M \partial^K  X^M\;. 
\ee

Before we continue we introduce an index-free notation that shall be useful later. 
In the following we will write all tensor equations `invariantly' by 
introducing an arbitrary basis $\{Z_{A}\}$, $A=1,\ldots, 2D$, 
that will later be specified to a coordinate basis, $Z_{M} = \partial_M$,  
or to a frame basis with additional constraints, $Z_{A}=E_{A}$, 
and accordingly the index $A$ will acquire different interpretations. 
For the moment, however, we keep the basis completely generic. 
With respect to this basis $\{Z_{A}\}$ and its dual $\{Z^{A}\}$ we 
have for the components of the metric (\ref{scalarpr}) 
 \be\label{frameMEtric}
  \vev{Z_A , Z^B} \ = \ \delta_A{}^B\;, \qquad 
  \vev{Z_A , Z_B} \ \equiv \ {\cal G}_{AB}\;.
 \ee
We stress that the metric ${\cal G}_{AB}$ will in general be $X$-dependent 
and it reduces to the constant $O(D,D)$ metric only for the coordinate basis. 
Under a change of basis $Z\rightarrow \tilde{Z}$ we have 
 \be\label{basischange}
  \tilde{Z}_{A} \ = \ \Lambda_{A}{}^{B} Z_{B}\;, \qquad \tilde{Z}^{A} \ = \ 
  (\Lambda^{-1})_{B}{}^{A} Z^{B}\;,
 \ee
where $\Lambda$ is an arbitrary, generally $X$-dependent, $GL(2D)$ matrix.   
Note that this transformation leaves the natural pairing in the first equation of (\ref{frameMEtric}) invariant. 
Accordingly, all definitions to be discussed in the following will be manifestly invariant  under a change 
of basis and in this sense be basis independent. 
For instance, 
with respect to this general basis we can then write for the D-bracket~(\ref{firstDbracket})  
\be\label{covDbracket}
  [X, Y]_D  \ =  \   [X, Y]_C  +   \frac{1}{2}\sum_A  (Z_A \, \vev{X, Y}) Z^A\;, 
 \ee
which is manifestly invariant under (\ref{basischange}) and reduces to (\ref{firstDbracket})
when using a coordinate basis.   
The lack of antisymmetry of the D-bracket is then expressed by 
\be
\label{indfreed}
[X, Y]_D  \ =  \  
- [Y, X]_D  +  \,\sum_A  (Z_A \, \vev{X, Y}) Z^A \;,   
\ee 
or with the help of the inner product as 
\be
\label{indfreed-vm}
\langle [X, Y]_D\,, Z \rangle   \ =  \  
- \langle  [Y, X]_D\,, Z \rangle   +  Z \,  \vev{X, Y} \;.  
\ee

In the following we will use the Einstein summation convention also for 
basis indices $A,B,\ldots$ and define for the gradient vector acting on a  general
function $f$,
\be
\label{gradient}  
\nabla f \ = \  \vec{\partial} f  \ = \   (Z_A \, f ) Z^A \,.
\ee

We close this section by  collecting some further properties of the C- and D-brackets. 
Just like 
the C-bracket in (\ref{confunction}), 
the D-bracket acts on scalars as the Lie bracket
\be
[X, Y]_D  f  \ = \   [X, Y] \,f  \,.
\ee
Moreover, the D-bracket satisfies the (modified) Jacobi identity 
 \be\label{DorfmanJacobi}
  \big[ X,\big[ Y,Z \big]_{ D}\big]_{ D}- \big[\big[ X,Y \big]_{ D}, Z \big]_{ D}
  - \big[ Y,\big[ X,Z \big]_{ D}\big]_{ D} \ = \ 0\;, 
 \ee
while for the C-bracket, we have the C-Jacobiator $J_C$:
 \be
 \label{c-jacobiator}
 J_C  (X, Y, Z) \ \equiv  \ \sum_{X, Y, Z}^{cyc}  [ \,[ X, Y]_C \,, Z ]_C
  \ = \   {1\over 6} \,\vec{\partial} \sum_{X, Y, Z}^{cyc}  \vev{  [X, Y]_C ,  Z}\;, 
 \ee
using eqn.\,(8.29) from \cite{Hull:2009zb}. Here, the sum denotes the 
cyclic sum with unit strength (i.e., three terms with coefficient one).

According to (\ref{Ddef}) the generalized Lie derivative $\widehat{\cal L}_{X}$ acts on generalized vectors 
via the D-bracket.  
Since it also  leaves the $O(D,D)$ invariant 
metric invariant and $X \vev{Y, Z} = \widehat{\cal L}_X \vev{Y, Z} $   
we have for any vector fields $X,Y,Z$  
 \be
 \label{derdorf}
  X\,\vev{Y,Z} \ = \ \vev{[X,Y]_{ D},Z}+\vev{Y,[X,Z]_{ D}}\;.
 \ee
In terms of the C-bracket it then follows from (\ref{covDbracket})  that 
 \be
 \label{c-bracket-inner}
  X\,\vev{Y,Z} \ = \ \vev{[X,Y]_{ C},Z}+\vev{Y,[X,Z]_{ C}}
  + {1\over 2} Z \, \vev{X, Y}   + {1\over 2}  Y\, \vev{X, Z} \;, 
 \ee
which will be useful below.

\subsection{Finite gauge transformations and invariant tensors}\label{secfinite}
Let us briefly recall the invariant `index-free' definition of tensors. A tensor 
is a multi-linear map from vectors and their duals to a function (scalar). Since by means of the metric (\ref{scalarpr}) 
we can always identify a dual vector with a vector, in the following we will 
restrict ourselves to multi-linear maps of vectors only. For a tensor $T$ of rank $n$ we can then scale  out 
a function multiplying any of the $n$ vector entries, i.e., 
 \be
  T(A_1,\ldots, f A_p, \ldots,A_n) \ = \ f T(A_1,\ldots,  A_p, \ldots,A_n)\;,
 \ee
and similarly for all other arguments.  The usual `component' form  of a tensor (with `curved indices'
in physicists notation) is then obtained by evaluating the tensor with respect to the coordinate basis $\partial_M$, 
 \be\label{comp1}
  T_{M_1\ldots M_n} \ = \ T(\partial_{M_1},\ldots, \partial_{M_n})\;.
 \ee
By its multi-linearity, the action of $T$ on arbitrary vectors can be written in terms of 
components as follows 
 \be\label{comp2}
  T(A_1,\ldots, A_n) \ = \ A_1^{M_1}\cdots A_n^{M_n}\,T(\partial_{M_1},\ldots, \partial_{M_n})
  \ = \ A_1^{M_1}\cdots A_n^{M_n}\,T_{M_1\ldots M_n}\;.
 \ee
  
Next we will show that a tensor thus defined coincides with a `generalized tensor' in the nomenclature  
of \cite{Hohm:2012gk}. There we introduced `generalized coordinate' transformations $X\rightarrow X'$ 
and defined a generalized vector $A_{M}$ as transforming according to 
 \be\label{finiteFIntro}
   A_{M}^{\prime}(X^{\prime}) \ = \ {\cal F}_M{}^N   A_N(X)\,,
   \ee
where the matrix ${\cal F}$ is defined by
   \be\label{calFdefIntro} \phantom{\Biggl(}
 {\cal F}_M{}^N\ \equiv \   {1\over 2} 
  \Bigl(   \frac{\partial X^{P}}{\partial X^{\prime M}}\,
  \frac{\partial X^{\prime}_P}{\partial X_{N}}
  + \frac{\partial X'_{ M}}{\partial X_P}\,
  \frac{\partial X^N}{\partial X^{\prime P}}\Bigr)
  \,, 
 \ee
and the indices on coordinates are raised and lowered with $\eta_{MN}$. 
We take this to be the definition of a generalized vector since, as mentioned above, 
currently we do not know of an invariant `intrinsic' definition. 
An arbitrary $O(D,D)$ tensor $T$ transforms as 
 \be\label{gentensor}
  T'_{M_1\ldots M_n}(X') \ = \ {\cal F}_{M_1}{}^{N_1}\cdots {\cal F}_{M_n}{}^{N_n}\, 
  T_{N_1\ldots N_n}(X)\;.
 \ee 
For 
an infinitesimal transformation $X'=X-\xi(X)$, we can expand  ${\cal F}$ 
to first order in $\xi$
and confirm that this transformation leads to the generalized Lie derivative (\ref{genLieder}). 
Thus, our current definition of a generalized tensor is the proper extension of our previous 
infinitesimal definition.  
 
The matrix ${\cal F}$ has various useful properties that are not manifest from 
its definition but that have been proved in  \cite{Hohm:2012gk}. First, due to the 
strong constraint (\ref{constraint}), (\ref{sconstraint}), 
  a transformation by ${\cal F}$ is actually 
compatible with the transformation of $\partial_M$ according to the chain rule, 
 \be\label{newchain}
  \partial_M' \ = \ {\cal F}_{M}{}^{N}\partial_N\;.
 \ee   
Second,  ${\cal F}\in O(D,D)$, i.e., 
a transformation by ${\cal F}$ is compatible with the metric (\ref{scalarpr}), 
 \be
  \langle V,W\rangle \ = \ \langle {\cal F} V, {\cal F}W\rangle\;,
 \ee
which implies in components 
 \be\label{inveta}
  \eta^{MN} \ = \ \eta^{KL}{\cal F}_{K}{}^{M}{\cal F}_{L}{}^{N}\;.
 \ee   
With these relations it then immediately follows that a tensor defined abstractly leads to a component tensor  
that is a generalized tensor in the sense of (\ref{gentensor}): from the multi-linearity of a tensor 
together with (\ref{comp1}) and  (\ref{newchain}) we infer 
 \be
 \begin{split}
  T'_{M_1\ldots M_n} \ &= \ T(\partial_{M_1}',\ldots, \partial_{M_n}') \ = \ 
  T( {\cal F}_{M_1}{}^{N_1} \partial_{N_1},\ldots, {\cal F}_{M_n}{}^{N_n}\partial_{N_n})\\
  \ &= \ 
  {\cal F}_{M_1}{}^{N_1}\cdots 
  {\cal F}_{M_n}{}^{N_n}\,T( \partial_{N_1},\ldots, \partial_{N_n}) 
  \ = \ {\cal F}_{M_1}{}^{N_1}\cdots 
  {\cal F}_{M_n}{}^{N_n}\,T_{N_1\ldots N_n}\;.
 \end{split}
 \ee 
Alternatively, using (\ref{inveta}) and the transformation (\ref{finiteFIntro}) of a generalized vector,  
we can read off the transformation of a generalized tensor from the right-hand side of (\ref{comp2}), 
using that the geometric (invariant) left-hand side is unchanged under coordinate transformations. 
In total, we can introduce generalized tensors in an `intrinsic' index-free fashion, if we take 
generalized vectors as given. 

We close this section by showing that the C- and D-brackets introduced above are well-defined 
brackets of generalized vectors, i.e., given two generalized vectors they produce a generalized 
vector. Since there is no intrinsic definition of generalized vectors we have 
to verify that, say, the C-bracket transforms correctly under generalized coordinate transformations. 
To this end it is convenient to employ an alternative form of the finite gauge transformations, 
which is simply given by the exponential of the generalized Lie derivative, 
 \be
  A_M'(X) \ = \ (\exp \widehat{\cal L}_{\xi})\, A_{M}(X)\;.   
 \ee
It has been shown in \cite{Hohm:2012gk} that, at least up to and including quartic order in $\xi$, 
this agrees with (\ref{finiteFIntro}) for a suitably defined generalized coordinate transformation 
$X'^M = X^M-\xi^M(X)+{\cal O}(\xi^2) $. 
This form of the finite gauge transformations is more convenient due to the following 
invariance property of the C-bracket \cite{Hohm:2010xe}
 \be
  \widehat{\cal L}_{\xi}\big[X,Y\big]_{C} \ = \ \big[\widehat{\cal L}_{\xi}X,Y\big]_{C}+\big[X,\widehat{\cal L}_{\xi}Y\big]_{C}\;. 
 \ee
Indeed, it is then easy to see that 
 \be
  e^{\widehat{\cal L}_{\xi}}\big[X,Y\big]_{C} \ = \ \big[e^{\widehat{\cal L}_{\xi}}X,e^{\widehat{\cal L}_{\xi}}Y\big]_{C}\;, 
 \ee
so that the C-bracket indeed transforms as a generalized vector. It is also easy to see that 
the D-bracket transforms as a generalized vector.
We first note from (\ref{firstDbracket}) that the 
D-bracket differs from the C-bracket by the partial derivative of a scalar. From 
(\ref{newchain}) it follows that the partial derivative of a scalar transforms like a vector. 
Thus, the D-bracket transforms also like a vector. 
In the next section we turn to the definition of Riemann and torsion tensors, which will be 
tensors in the invariant sense recalled above and which will be 
written in terms of the C- and D-bracket. From the foregoing discussion it is then 
clear that these tensors are generalized tensors in the sense above and, therefore, 
that all actions build with these curvatures are invariant under finite gauge transformations.

\section{Invariant geometry of double field theory}
In this section we define the Riemann and torsion tensors
appearing in double field theory along the lines of an invariant approach reviewed above. 
This means that we may freely choose to 
evaluate these objects with respect to a coordinate (holonomic) basis or 
with respect to an arbitrary (anholonomic) frame. 
This invariant formulation therefore provides  
a unified description of a `metric-like' and `frame-like' formalism.

\subsection{Covariant derivatives} 
We start by introducing covariant derivatives or connections in the usual invariant fashion.  
One defines a connection $\nabla$ as 
a bilinear operator that, given two vector fields $X, Y $, provides a third:
\be
(X, Y) ~\to ~  \nabla_X Y  \,, 
\ee
where bilinear means that for constants $a, b$ and functions $f, g$ on the manifold we have
\be
\begin{split}
\nabla_X ( aY_1 + bY_2) \ = & \  a\nabla_X Y_1 + b\nabla_X Y_2\,, \\
\nabla_{f X_1+ g X_2}  Y\ = & \  f\, \nabla_{X_1} Y + g\,  \nabla_{X_2}  Y\,. 
\end{split}
\ee
Moreover we must also have 
\be
\label{scaling}   
\nabla_X \,fY  \ = \    X(f) Y  +  f \nabla_X Y \,.
\ee
We also write 
\be
\nabla_X f  \ \equiv  \  X(f) \ = \ X^M \partial_M f \,.
\ee
In that way we can make 
(\ref{scaling}) look like a derivation:
\be
\nabla_X \,fY  \ = \    (\nabla_X f) Y  +  f \nabla_X Y \,.
\ee
We require that the metric is compatible with the connection $\nabla$, 
 \be\label{metriccomp}
  \nabla_{X}\vev{Y,Z} \ = \ X\, \vev{Y,Z} \ = \ \vev{\nabla_{X}Y,Z}+\vev{Y,\nabla_{X}Z}\;.
 \ee

Next, we extend the covariant derivative to arbitrary tensors. 
Consider a $p$-tensor $K$ that, given $p$ vector entries, 
gives a function (number) $K(X_1, X_2, \cdots ,X_p )$.  Its covariant derivative
$\nabla K$ is defined as a $p+1$ tensor:
\be
\nabla K (X_1, X_2, \cdots ,X_p , W) \ \equiv  \  \nabla_W 
K (X_1, X_2, \cdots ,X_p)\,, 
\ee
where the $\nabla_W$ action on $K$ gives a $p$-tensor defined by
\be\label{covderaction}
\begin{split}
(\nabla_W  K) (X_1, X_2, \ldots , X_p) \ \ \equiv \  \ &  ~~~W \cdot K (X_1, X_2, \ldots , X_p)\\
& -  K \,(\,\nabla_W X_1 , X_2, \ldots , X_p)  \\
& -  K \,(\, X_1 , \nabla_W X_2 ,\ldots , X_p) \\
& ~~~~~~\vdots  \\
& -    K \,(\, X_1 , X_2, \ldots , \nabla_WX_p) \,.\\
\end{split}
\ee
A useful subcase arises when the $p$-tensor $K$ is defined via
an inner product and a vector ${\cal K}$ that is a function of
$p-1$ vectors:
\be
K(X_1, X_2, \ldots, X_{p-1}\,, X_p) \ = \ \vev{{\cal K} (X_1, X_2,\ldots, X_{p-1}) \,, X_p}  \;. 
\ee
Then
\be
\begin{split}
\nabla_W K(X_1, X_2, \ldots, X_{p-1}\,, X_p)
 \ = & \ ~ ~  \vev{\nabla_W {\cal K} (X_1, X_2,\ldots, X_{p-1}) \,, X_p}  \\
 & -  \vev{{\cal K} \,(\,\nabla_W X_1 , X_2, \ldots X_{p-1}) , X_p}  \\
& ~~~~~~\vdots  \\
& -   \vev{  {\cal K} \,(\, X_1 , X_2,\, \ldots , \nabla_WX_{p-1}), X_p} \,, 
\end{split}
\ee
where we used (\ref{metriccomp}) to cancel two terms with $\nabla_W X_p$ in the inner product. 
We see that the inner product remains as a spectator.

The Lie bracket (\ref{usualLie})  of two vectors is another vector
(but \textit{not} a generalized vector), that can be defined by the action on
a function as follows
\be
 ~\nabla_{[X,Y]}f \ =  \  [X,Y] (f) \  \equiv  \  X (Y (f)) -  Y (X (f)) \,.
\ee
Using our nabla notation for the action on functions we can write the above
as
\be
\nabla_{[X,Y]} f \  =   \  \nabla_X (\nabla_Y f) -  \nabla_Y (\nabla_X f) \ = \  [\nabla_X\,, \nabla_Y] f \,, 
\ee
 and therefore for functions
\be
\label{comscal}
   [\nabla_X\,, \nabla_Y] \,f  
 \ = \ \nabla_{[X,Y]} f\,.  \ee
We also note the property
\be
\label{ten-lie}
 [fX, Y] \ = \   f [ X, Y]  -   (\nabla_Y f) X  \,,   
\ee
which by the various linearity and scaling properties implies that:
\be
\label{ten-lie2}
\begin{split}
 ~\nabla_{[fX, Y]} \ = & \   f\, \nabla_{[ X, Y]}  
 -  ( \nabla_Yf) \nabla_X  \,. ~  \\
\nabla_{[X, gY]} \ = & \   g\, \nabla_{[ X, Y]}  +  ( \nabla_Xg) \nabla_Y  \,. 
\end{split}
\ee

Next, we compute the scaling properties of the C- and D-brackets under $X\rightarrow f X$, 
which will be needed later to verify tensor properties. We compute with the help of (\ref{cbrakc})
\be
\label{cour-scal}
\vev{[fX, Y]_C \,, Z }  \ = \ f \vev{[X,Y]_C\,, Z} 
-  (\nabla_Y f) \vev{X, Z}  
+ {1\over 2} (\nabla_Z f) \vev{X, Y}  \;. 
\ee
We also have the following scaling:
\be
\label{cbrscal}
\nabla_{[fX, Y]_C}   \ = \ f  \nabla_{[X, Y]_C}   \ - \ (\nabla_Y f)  \nabla_X \,    + {1\over 2}  \vev{X, Y}  \nabla_{\nabla f}  \,,
\ee
where again $\nabla f$ is the vector with components $\partial^M f$ so that
\be
\nabla_{\nabla f}  \ = \ ( \partial^M f\,)  \nabla_{\partial_M}   \;.   
\ee  
For the D-bracket we have
\be
\label{d-cov-scal}
\begin{split}
\nabla_{[fX, Y]_D}   \ = &\  \ f  \nabla_{[X, Y]_D}   \ - \ (\nabla_Y f)  \nabla_X \,    +  \vev{X, Y}  \nabla_{\nabla f}  \,,\\
\nabla_{[X, gY]_D}   \ = &\  \ g  \nabla_{[X, Y]_D}   \ + \ (\nabla_X g)  \nabla_Y \,     \,.\\
\end{split}
\ee
As we can see, compared to the Lie bracket (\ref{ten-lie}), only the 
first input of the D-bracket 
scales differently.  
 We also have
\be
\label{dorfsf-scal}
\begin{split}
\vev{[fX, Y]_D \,, Z }  \ = & \ f \vev{[X,Y]_D\,, Z} 
-  (\nabla_Y f) \vev{X, Z}  
+  (\nabla_Z f) \vev{X, Y}\;,   \\
\vev{[X, gY]_D \,, Z }  \ = & \ g \vev{[X,Y]_D\,, Z} 
+  (\nabla_X g) \vev{Y, Z}  \;. 
\end{split}
\ee
Another rescaling property is
\be
\label{scalednablas}
\begin{split}
\nabla_{[X, Y]_C}  f Z \ = & \  f \nabla_{[X, Y]_C}   Z  + (\nabla_{[X, Y]} f) Z \,,\\
\nabla_{[X, Y]_D}  f Z \ = & \  f \nabla_{[X, Y]_D}   Z  + (\nabla_{[X, Y]} f) Z \;, 
\end{split}
\ee
which scale like the Lie bracket, since the action of the C- and D-bracket on 
functions is the same as that of the Lie bracket.

\subsection{Generalized torsion}
We now aim to define a generalized torsion tensor. 
Before doing so let us first recall the 
torsion tensor of conventional differential geometry.  Using lowercase
letters to denote vector fields    
\be
\label{def-tor-gen}
T (x, y) \ \equiv \  \nabla_x y  - \nabla_y x  - {[x\,, y]}  \,.
\ee
Here $ T(x,y)$  
is itself a vector.  We will find it convenient to define a torsion
tensor with three inputs using the inner product:
\be
\label{def-tor-gen-3inputs}
T (x, y, z) \ \equiv  \ \vev{ T(x, y) , z}  \ = \ \vev{ \nabla_x y  - \nabla_y x  - {[x\,, y]}, z}   \,.
\ee
It is straightforward to prove the scaling (tensorial) property $T(fx, y) = f T(x,y)$, 
\be
\begin{split}
T (fx, y) \ = & \ \nabla_{fx} y  - \nabla_y fx  - 
[fx\,, y] \\   
= & \ f\nabla_{x} y  - ( f\nabla_y x  + (\nabla_yf) x ) - ( f [x\,, y]
- (\nabla_yf) x)   \\
= & \  f \,T(x,y) \,.
\end{split}
\ee

Let us now generalize the torsion tensor (\ref{def-tor-gen}).  We want to change
the bracket to the C-bracket, because otherwise we do not get generalized vectors.
But that actually ruins the scaling property.  This can be fixed with extra terms:
\be
\label{defgentor}   
 {\cal T}_0 (X, Y, Z) \ 
\equiv  \    \vev{ \nabla_X Y - \nabla_Y X - \, [X, Y]_C\, , \, Z } -{1\over 2} \vev{X, \nabla_Z Y}
+{1\over 2} \vev{Y, \nabla_Z X}  
 \,. 
\ee
Note the second and third terms on the right-hand side use the $Z$ entry in
a nontrivial way.  The scaling $Z \to f Z$ works manifestly on both sides
of the equation.
The scaling $X \to f X$ requires the extra terms and uses (\ref{cour-scal})
to show that it works.   So this defines a generalized torsion
tensor.  
Formally, this definition agrees with that given 
by Gualtieri, see Def.~3 in \cite{Gualtieri:2007bq}. 

Let us now determine the component form of the torsion tensor in a 
coordinate basis, $Z_{M}=\partial_{M}$. With respect to this basis 
we define (Christoffel) connection components via 
 \be\label{Christoffel-like}
  \nabla_{\partial_{M}}(\partial_{N})\ = \ \Gamma_{MN}{}^{K}\partial_{K}\;.
 \ee
A calculation then shows  
\be\label{torsion1comp}
{\cal T}_0(\partial_{M},\partial_{N},\partial_{K}) \ = \ ({\cal T}_{0})_{MNK}  \ = \ \Gamma_{MNK}  - \Gamma_{NMK} -{1\over 2} \Gamma_{KNM}
+{1\over 2} \Gamma_{KMN}\,, 
\ee
in agreement with the definition given in \cite{Hohm:2011si}. 

The above torsion is based on the C-bracket. We can give an alternative, simpler torsion tensor 
${\cal T}$ using the D-bracket: 
\be
\label{defgentor}
 {\cal T} (X, Y, Z) \ 
\equiv  \    \vev{ \nabla_X Y - \nabla_Y X- [X, Y]_D\, , \,Z } + \vev{Y, \nabla_Z X}  \,. 
\ee
A quick computation using (\ref{dorfsf-scal})  establishes the tensorial nature of this definition. 
A calculation shows that
\be
\label{comp-torsion}
{\cal T} _{MNQ}  \ = \ \Gamma_{MNQ}  - \Gamma_{NMQ} + \Gamma_{QMN}\,.
\ee
To see that this agrees with (\ref{torsion1comp}) we 
use that the metricity condition (\ref{metriccomp}) specialized to a coordinate basis:  
 \be
  0 \ = \ 
  \partial_{M}\eta_{NK} 
  \ = \ \vev{\nabla_{\partial_{M}}(\partial_{N}),\partial_{K}}+\vev{\partial_{N},\nabla_{\partial_{M}}(\partial_{K})}
  \ = \ \Gamma_{MNK} + \Gamma_{MKN}\;, 
 \ee
from which we infer the antisymmetry of  $\Gamma_{MNK}$ in its last two indices.  
We then see that
\be
\label{comp-torsion99}
{\cal T} _{MNQ}  \ = \ \Gamma_{MNQ}  + \Gamma_{NQM} + \Gamma_{QMN}\,, 
\ee
and therefore ${\cal T}$ is  cyclic and  totally antisymmetric 
and agrees with (\ref{torsion1comp}). 
 These are not so obvious from the
geometrical definition, but 
 short calculations using (\ref{indfreed-vm}) and (\ref{derdorf})  show that 
\be
\begin{split}
 {\cal T} (Y, X, Z) \ = \ & \ - {\cal T} (X, Y, Z)  + \vev{  \nabla_Z X, Y } + \vev{ X, \nabla_Z Y} - Z \vev{X, Y}\;,  \\
  {\cal T} (X, Z, Y) \ = \ & \ - {\cal T} (X, Y, Z)  + \vev{  \nabla_X Y, Z } + \vev{ Y, \nabla_X Z} - X \vev{Y, Z} \;. 
\end{split}
\ee
Indeed, when the connection is compatible with the metric we get  
\be
 {\cal T} (X, Y, Z) \ = \ - {\cal T} (Y, X, Z) \ = \ -  {\cal T} (X, Z, Y)\,.
\ee

For a further rewriting, we return to a general basis and use that for any vector $X$ we have
the expansion $X =  X^M Z_M$ so that 
\be\label{AnablaB}
\begin{split}
\vev{Y, \nabla_X Z} \ = \ & \    \vev{Y, \nabla_{Z_M} Z} \, X^M
\ = \   \vev{Y, \nabla_{Z_M} Z} \, \vev{Z^M, X} \\
 = \ & \  \bigl\langle ~  \vev{Y, \nabla_{Z_M} Z} Z^M\,, \,X\,
 \bigr\rangle \;. 
\end{split}
\ee
With this result we can then write (\ref{defgentor}) as 
\be
{\cal T} (X, Y, Z) \ = \ \vev{  {\cal T}^{\uparrow} (X, Y) , Z } \,, 
\ee
with
\be\label{torsionup} 
{\cal T}^{\uparrow} (X, Y) \ = \ \nabla_X Y - \nabla_Y X - [X, Y]_D  
+  \vev{Y, \nabla_{Z_A} X} Z^A\,. 
\ee
This is the invariant form of the torsion tensor with one index raised with the $O(D,D)$ invariant metric.

\subsection{Generalized Riemann tensor} 

We now attempt to define a generalized  Riemann tensor. 
As a first try, we take   
the standard invariant definition (\ref{theoriginal}) of 
Riemannian geometry and  replace the Lie bracket by either the Courant bracket 
or Dorfman bracket: 
\be
\label{test0}
\begin{split} 
R (X, Y) Z  \ \equiv  & \ \   (\nabla_X \nabla_Y -  \nabla_Y \nabla_X -  \nabla_{[X, Y]_C}) Z       ~~? \;, \\
R (X, Y) Z  \ \equiv  & \ \   (\nabla_X \nabla_Y -  \nabla_Y \nabla_X
-  \nabla_{[X, Y]_D} ) Z        ~~? \; 
\end{split}
\ee
Recalling  that  
both $\nabla_{[X, Y]_C}  Z$ and $\nabla_{[X, Y]_D} Z$
rescale like  $\nabla_{[X, Y]} Z$ under $Z \to f Z$  (see (\ref{scalednablas})), we conclude
 that these rescale correctly under $Z \to f Z$.  
Neither, however, rescales correctly under $X \to f X$, as can be seen from 
inspection of (\ref{cbrscal}) and (\ref{d-cov-scal}).  
In addition, for $Y \to gY$, the first (Courant)  has anomalous rescaling
while the second (Dorfman) has correct rescaling. 
Given the extra simplicity of the Dorfman bracket we will now use it to attempt
the full construction of a tensor ${\cal R}$. 
For the comparison below with formulas in the literature it is, however, convenient 
to give a name to the (non-tensorial) object defined with the C-bracket, 
 \be
\label{test}  
R (X, Y, Z,W)  \ \equiv    \   \vev{ \, (\nabla_X \nabla_Y 
-  \nabla_Y \nabla_X-  \nabla_{[X, Y]_C}) Z \,  ,\, W}    \;. \\
\ee

We begin now the construction of the covariant curvature. 
Using the metric to have extra
flexibility in writing terms,  we begin with   
\be  
{\cal R} (X, Y, Z, W)  \ \equiv   ~\vev{(\nabla_X \nabla_Y -  \nabla_Y \nabla_X - \nabla_{[X, Y]_D} ) Z \,, W} + \ldots \;, 
\ee
where the dots denote terms to be added and so far the extra
input $W$ has played no role.  Since we use Dorfman  this
definition is not $X,Y$ antisymmetric. Clearly there is no 
rescaling problem with $W$.  Again, as discussed above, there is no rescaling problem
for $Z$. There is no rescaling problem for 
$Y$, since the Dorfman bracket transforms like the Lie bracket
for scalings of the second argument, as can be seen by comparing (\ref{dorfsf-scal}) with (\ref{ten-lie}), 
and thus the proof of scaling
for the ordinary curvature tensor suffices.  The only problem is the scaling of $X$.

Let us compute the anomalous term for $X$ scaling in the above curvature, 
denoted by Anom, by which we mean the anamalous terms \textit{beyond} those of 
the Lie bracket. This 
requires using   the
first equation of (\ref{d-cov-scal}), which gives
\be
 \hbox{Anom~} \nabla_{[fX, Y]_D} \ = \   \vev{X, Y} \, \nabla_{\nabla f} \,, 
\ee
so that in the above curvature we get an anomalous term 
\be
\label{tbcan}
\hbox{Anom~} (-  \vev{\nabla_{[fX, Y]_D} Z, W})  \ = \   -\, \vev{X, Y} \, \vev{   \nabla_{\nabla f} Z, W} \,.
\ee
To cancel it we must add some term to the definition of the curvature.
We want a term that is a generalized scalar with problematic $X$ scaling and good $Y$ scaling
(and ideally good $Z$ and $W$ scaling).  We can come quite close
to this by adding the term 
\be
\Delta {\cal R} \ = \  \vev{Y, \nabla_{Z_A} X}  \, \vev{W, \nabla_{Z^A}  Z }  \;.
\ee
In a coordinate basis this term would read
$Y_K (\nabla^Q X)^K  W_N (\nabla_Q Z)^N$.  
As desired, 
$\Delta {\cal R}$  does not scale anomalously for $Y$.  
For $X \to f X$
we have an extra anomalous term
\be  
\vev{Y,  X}\,  (Z_A\,  f)\, \vev{W, \nabla_{Z^A} Z }\ = \ 
\vev{X,  Y} \, \vev{ \nabla_{  (Z_A\cdot  f)Z^A} Z, W }\ = \ 
\vev{X,  Y} \, \vev{ \nabla_{\nabla f} Z, W }\;,
\ee
using (\ref{gradient}). The above term cancels precisely (\ref{tbcan}). 
Next, 
$\Delta {\cal R}$ has good scaling with $W$ but now the $Z$ 
scaling has been compromised.   The new 
term $\Delta' {\cal R}$ 
required to  
cancel the $Z$ scaling of $\Delta {\cal R}$  is   
\be
\Delta'{\cal R} \ = \  -  \vev{\nabla_{[Z, W]_D} X, Y}\;, 
\ee
as can be seen with the first equation in (\ref{d-cov-scal}).  
But this time the conventional $Z$ scaling and $W$ scalings 
are ruined, since we do not have the extra terms in the curvature,
so we finally take
\be
\label{curvaturehere_it_is}  
\begin{split}
\phantom{\Bigl(}~~{\cal R} (X, Y, Z, W)  \ \equiv & \ \ \ ~ ~\vev{(\nabla_X \nabla_Y -  \nabla_Y \nabla_X- \nabla_{[X, Y]_D} ) Z \,, W} \\
& \ \  + \vev{(\nabla_Z \nabla_W -  \nabla_W \nabla_Z - \nabla_{[Z, W]_D}) X \,, Y}  ~~\\
& \ \  +  \vev{Y, \nabla_{Z_A} X}  \, \vev{W, \nabla_{Z^A} Z }\,. \phantom{\Bigl(}
\end{split}
\ee
One can now verify that all scalings work out so that ${\cal R}$ as defined here is a tensor.

By definition, it is manifest   
that ${\cal R}$ is symmetric under the exchange of the first two inputs with the last
two inputs: 
\be\label{exchangesym}
{\cal R} (X, Y, Z, W) \ = \ {\cal R} (Z, W,X, Y)\,.
\ee
In ordinary geometry this does not hold unless the torsion vanishes.
The antisymmetry  
 in the first or second pair of arguments is not too hard to show.  Consider
the exchange of the first two arguments, 
\be
\label{vmvgsct}
\begin{split}
{\cal R} (Y, X, Z, W)  \ \equiv & \ \ \  ~\vev{(\nabla_Y \nabla_X -  \nabla_X \nabla_Y - \nabla_{[Y, X]_D} ) Z \,, W} \\
& \ \  + \vev{(\nabla_Z \nabla_W -  \nabla_W \nabla_Z - \nabla_{[Z, W]_D}) Y \,, X}  ~~\\  
& \ \  +  \vev{X, \nabla_{Z_A} Y}  \, \vev{W, \nabla_{Z^A} Z }\;. 
\end{split}
\ee
To deal with the second line we 
recall (\ref{comscal}) 
 and that acting on functions the Lie and D-brackets
coincide: 
\be
\label{adqui99}
(\nabla_Z \nabla_W  - \nabla_W \nabla_Z) \vev{X, Y}  \ = \  \nabla_{[Z, W]_D}
\vev{X, Y} \;. 
\ee
Letting the derivatives act we find
\be
\label{adqui}
 \vev{(\nabla_Z \nabla_W  - \nabla_W \nabla_Z-\nabla_{[Z, W]_D})X, Y} 
+  \vev{X, (\nabla_Z \nabla_W  - \nabla_W \nabla_Z-\nabla_{[Z, W]_D})Y}  \ = \  0 \;. 
\ee
Using this in the second line of (\ref{vmvgsct}) and using (\ref{indfreed}) 
in the first line  
we find 
\be
\begin{split}
{\cal R} (Y, X, Z, W)  \ \equiv & \ \ \  -\vev{(\nabla_X \nabla_Y -  \nabla_Y \nabla_X +  \nabla_{-[X, Y]_D 
+  (Z_A \vev{X, Y}  Z^A } ) 
Z \,, W} \\
& \ \ -  \vev{(\nabla_Z \nabla_W  - \nabla_W \nabla_Z-\nabla_{[Z, W]_D})X, Y}  ~~\\
& \ \  +  \vev{X, \nabla_{Z_A} Y}  \, \vev{W, \nabla_{Z^A} Z }\;. 
\end{split}
\ee
This gives 
\be
\begin{split}
{\cal R} (Y, X, Z, W)  \ \equiv & \ \ \  
-\vev{
(\nabla_X \nabla_Y -  \nabla_Y \nabla_X -  \nabla_{[X, Y]_D} ) Z, W}  
-  (Z_A \, \vev{X, Y}) \vev{W , \nabla_{Z^A} Z} \\
& \ \ -  \vev{(\nabla_Z \nabla_W  - \nabla_W \nabla_Z-\nabla_{[Z, W]_D})X, Y}  ~~\\
& \ \  +  \vev{X, \nabla_{Z_A} Y}  \, \vev{W, \nabla_{Z^A} Z }\;. 
\end{split}
\ee
The last term on the first line combines with the last term of the
right-hand side to give  
\be
\begin{split}
{\cal R} (Y, X, Z, W)  \ \equiv & \ \ \  
-\vev{
(\nabla_X \nabla_Y -  \nabla_Y \nabla_X -  \nabla_{[X, Y]_D} ) Z, W}  
 \\
& \ \ -  \vev{(\nabla_Z \nabla_W  - \nabla_W \nabla_Z-\nabla_{[Z, W]_D})X, Y}  ~~\\
& \ \  -     \vev{Y, \nabla_{Z_A} X}  \, \vev{W, \nabla_{Z^A} Z }\;. \end{split}
\ee
This shows that, as claimed,   
\be
\label{exchsignvm}
{\cal R} (Y, X, Z, W)  \ = \ -  {\cal R} (X, Y, Z, W) \,.  
\ee

Next, we examine the component expansions.  We write, 
\be
{\cal R} (X, Y, Z, W)   \ = \  X^M Y^N \, Z^K W^L \, {\cal R}_{MNKL}\;. 
\ee
In a coordinate basis the last term in the curvature formula (\ref{curvaturehere_it_is})
reads   
\be
 (Y_N \nabla_{\partial^Q} X^N)  \, (W_K \nabla_{\partial_Q} Z^K)\,. 
\ee
By the scaling property we know that only the part
without derivatives on any of $X,Y,W$ and $Z$ contributes to the curvature components. 
We thus find
\be
X^M Y^N  \,\Gamma^Q{}_{MN}  ~ Z^K W^L \, \Gamma_{QKL} \;,   
\ee
giving 
the following contribution to ${\cal R}_{MNKL}$:
\be
 \Gamma^{Q}{}_{MN} \Gamma_{QKL} \;. 
\ee
Since the Dorfman bracket gives
exactly the same derivative-independent terms as the Lie bracket or the Courant bracket,  
we have proven that the geometric definition of ${\cal R}$ above
coincides with the one in~\cite{Hohm:2011si}:
\be\label{calRiemcomp}
 {\cal R}_{MNKL} \ = \   {R}_{MNKL}+  {R}_{KLMN}  +  \Gamma^{Q}{}_{MN} \Gamma_{QKL} \,, 
\ee
where the components of $R$ arise from (\ref{test}) and read 
 \be 
    R_{MNKL} \ = \ \partial_{M}\Gamma_{NKL}-\partial_{N}\Gamma_{MKL}
  +\Gamma_{MQL}\Gamma_{NK}{}^{Q}-\Gamma_{NQL}\Gamma_{MK}{}^{Q}\;.
 \ee

\section{Algebraic Bianchi identity}
Here we apply the above geometrical framework to prove an
 algebraic Bianchi identity 
that holds without imposing the constraint that the generalized
torsion vanishes.  
Specializing to a coordinate basis 
and setting the generalized  
torsion to zero, this leads to the algebraic Bianchi identity derived in \cite{Hohm:2011si}
for the component Riemann tensor.

\subsection{Invariant form of algebraic Bianchi identity}\label{algebraicSec1} 
The algebraic Bianchi identity that we will prove can be written as 
 \be\label{INVAlgBianchi}
  \sum_{W, X, Y, Z}^{\rm ant}\Big(3\,{\cal R}(W,X,Y,Z)-4\, \nabla_{W}{\cal T}(X,Y,Z)-
  3{\cal T}(W,X,{\cal T}^{\uparrow}(Y,Z))\Big) \ = \ 0\;,
 \ee
which holds for arbitrary vector fields $X,Y,Z$ and $W$, and 
the sum is to be interpreted as complete antisymmetrization over the four arguments. 
In this form it looks weaker than the algebraic Bianchi identity in conventional 
Riemannian geometry, for the latter involves only an antisymmetrization over 
three rather than four arguments.  It turns out, however, that due to the extra pair 
exchange 
symmetry (\ref{exchangesym})  of the generalized Riemann tensor (as compared to the conventional one)  
this form of the Bianchi identity is equivalent to a similar identity with antisymmetrization over 
three arguments only, as we will now show.

To this end we find it convenient to write (\ref{INVAlgBianchi}) with respect to the coordinate basis, 
which then reads   
  \be\label{finaltorsionidentity}
  3\, {\cal R}_{[MNKL]} \ = \ 4\nabla_{[M}{\cal T}_{NKL]} 
  +3{\cal T}_{[MN}{}^{Q}{\cal T}_{KL]Q}\;. 
 \ee
We will now show that this identity is equivalent to 
 \be\label{BianchiIDENTITy}
\begin{split}
\sum_{M,N,K}^{cyc} {\cal R}_{MNKL}  
 \ = \ -\,\nabla_L {\cal T}_{MNK}+ \sum_{M,N,K}^{cyc} 
 \Bigl( \nabla_M {\cal T}_{NKL}   + {\cal T}_{MN}{}^Q {\cal T}_{KLQ} \Bigr)\;, 
\end{split}
\ee
where the cyclic sum extends over three arguments.  
First, note that the cyclic sums actually create antisymmetry
in three indices.  For this recall that for a three-index tensor $S_{ABC}$ 
 that is 
antisymmetric in two 
indices, complete antisymmetrization is 
equivalent to the cyclic sum, 
\be
\label{cycant}
 S_{ABC} \ = \ - S_{BAC} \quad \Rightarrow \quad 
~\sum_{A,B, C}^{cyc}  S_{ABC}  \ = \  3 S_{[ABC]}\,.
\ee
Recalling that ${\cal T}$ is totally antisymmetric,  
we can use this to rewrite (\ref{BianchiIDENTITy}) as 
 \be
 \label{first-rw}
  3\,{\cal R}_{[MNK]L} \ = \ -\nabla_{L}{\cal T}_{MNK}
  + 3\,\nabla_{[M}{\cal T}_{NK]L}
  +3\,{\cal T}_{[MN}{}^{Q}{\cal T}_{K]LQ}\;.
 \ee
To proceed further consider the antisymmetrization identities
\be
\begin{split}
S_{[ABC]} \ = \  \  {1\over 3} \sum_{A,B,C}^{cyc}  S_{A [BC]} \,, ~~~
S_{[ABCD]} \ = \  \  {1\over 4} \sum_{A,B,C,D}^{\pm \,cyc}  S_{A [BCD]} \;, 
\end{split}
\ee
where the $\pm$ indicates that the cyclic sum alternates signs.  The 
second identity implies that 
\be
\nabla_{[M} {\cal T}_{NKL]}  \ = \  {1\over 4} \Bigl( 
\nabla_{M} {\cal T}_{[NKL]}  - \nabla_{N} {\cal T}_{[KLM]}  
+ \nabla_{K} {\cal T}_{[LMN]}  - \nabla_{L} {\cal T}_{[MNK]}  
\Bigr)\;. 
\ee
The total antisymmetry of  ${\cal T}$ allows us to delete the  
$[\ldots ]$ on its indices and thus we have 
\be
\nabla_{[M} {\cal T}_{NKL]}  \ = \  {1\over 4} \Bigl( 
\nabla_{M} {\cal T}_{NKL}  + \nabla_{N} {\cal T}_{KML}  
+ \nabla_{K} {\cal T}_{MNL}  + \nabla_{L} {\cal T}_{MKN}  
\Bigr)\;, 
\ee
where we used the antisymmetry of ${\cal T}$ to rearrange indices.  
This in turn can be rewritten as  
\be
\begin{split}
4\nabla_{[M} {\cal T}_{NKL]}  \ =   \ \  - \nabla_{L} {\cal T}_{MNK}  
+ \sum_{M,N,K}^{cyc} 
\nabla_{M} {\cal T}_{NKL}  
=   \ \  - \nabla_{L} {\cal T}_{MNK}  
+ 3\, \nabla_{[M} {\cal T}_{NK]L} \;, 
\end{split}
\ee
where we used (\ref{cycant}).  Thus, the combination of terms on the 
right-hand side of (\ref{first-rw}) 
with covariant derivatives on the torsion is actually totally antisymmetric
in four indices and so (\ref{first-rw}) becomes
 \be
 \label{first-rw2}
  3\,{\cal R}_{[MNK]L} \ = \ 4\,\nabla_{[M} {\cal T}_{NKL]} 
  +3\,{\cal T}_{[MN}{}^{Q}{\cal T}_{K]LQ}\;.
 \ee
This equation suggest that the left-hand side is actually antisymmetric
on the four indices, which we now show.  This fact is a consequence of the antisymmetry
in each pair and the {\em symmetry} under pair exchange,\footnote{Recall that 
this 
property of the generalized Riemann tensor 
holds even with torsion.  This is not the case for the conventional Riemann tensor.}
\be
\begin{split}
{\cal R}_{[MNK]L}  \ = \ &  \ {1\over 3} \Bigl(  
{\cal R}_{MNKL} + {\cal R}_{NKML}  + {\cal R}_{KMNL} \Bigr) \\
 = \ &  \ {1\over 3} \Bigl( - 
{\cal R}_{MNLK} - {\cal R}_{LMNK}  - {\cal R}_{NLMK} \Bigr) 
 \ = \  -  {\cal R}_{[MNL]K} \;. 
\end{split}
\ee
Being thus antisymmetric in its four indices the Riemann tensor satisfies
\be
{\cal R}_{[MNK]L}  \ = \ {\cal R}_{[[MNK]L]} \ = \ {\cal R}_{[MNKL]} \;. 
\ee
A completely analogous analysis shows that the combination 
${\cal T}_{[MN}{}^{Q}{\cal T}_{K]LQ}$ is also totally antisymmetric
in $M, N, K$, and $L$.  
Using this in (\ref{first-rw2}) finally proves that 
(\ref{finaltorsionidentity}) follows from (\ref{BianchiIDENTITy}), thus showing their equivalence.

We can understand group theoretically  
that for the generalized   
Riemann tensor antisymmetry in three indices 
implies antisymmetry in all four.  
Since ${\cal R}$ is antisymmetric in its first two and second two indices 
it lives in the tensor product     
  \bea
  {\small \yng(1,1)}\otimes {\small \yng(1,1)}\hspace{0.2em}
  \ = \  {\small \yng(1,1,1,1)} \oplus {\small \yng(2,1,1)} \oplus 
  {\small \yng(2,2)} \; ~. 
 \eea
By definition, however, 
${\cal R}$ has the exchange symmetry between the two 
index pairs independently of the torsion constraint. Therefore, ${\cal R}$ belongs only 
to the symmetric tensor product  
 \bea\label{irreducible}
  \left(\hspace{0.2em}{\small \yng(1,1)}\otimes {\small \yng(1,1)}\hspace{0.2em}\right)_{\rm sym}
  \ = \   {\small \yng(1,1,1,1)} \oplus 
  {\small \yng(2,2)} \; ~. 
 \eea
Antisymmetrization in three indices eliminates the window Young tableau ${\tiny \yng(2,2)}$, 
and therefore only the totally antisymmetric part survives. 
In the form (\ref{INVAlgBianchi}) the algebraic Bianchi identity is relatively easy to 
prove, as we do below.

\subsection{Invariant proof}
We will now give an invariant `index-free' proof of the algebraic Bianchi identity 
(\ref{INVAlgBianchi}) (and thus of its equivalent forms (\ref{BianchiIDENTITy}) and (\ref{first-rw2})). 
We first write (\ref{INVAlgBianchi}) as
 \be\label{INVAlgBianchi99}
 {\cal R}(W,X,Y,Z)\, -\, {4\over 3} (\nabla_{W}{\cal T})(X,Y,Z) \, - \, 
  {\cal T}(W,X,{\cal T}^{\uparrow}(Y,Z)) \ = \ 0 \ ,  
 \ee
where from now on we will leave the totally 
antisymmetric projection implicit. 
One important simplification due to the antisymmetry is that we can
replace D-brackets  for C-brackets, because these are precisely the antisymmetrization of D-brackets.  
Thus, we can replace $[X, Y]_D$ by $[X, Y]_C$  and, by linearity,  
 $\nabla_{[X, Y]_D}$
by $\nabla_{[X, Y]_C}$.  Therefore, we get 
\be
\label{curvred}
 {\cal R}(W,X,Y,Z) \ = \ 4 \vev{\nabla_W \nabla_X Y, Z}  - 2 \vev{
 \nabla_{[W, X]_C} Y , Z}  + 
  \vev{X , \nabla_{Z_A} W} \, \vev{Z,  \nabla_{Z^A} Y} \;.
 \ee
Let us consider the double torsion term in (\ref{INVAlgBianchi99}). We compute with (\ref{defgentor}) 
and (\ref{torsionup}) 
\be
\begin{split}
 {\cal T}(W,X,{\cal T}^{\uparrow}(Y,Z))\ = \ &  {\cal T} \bigl(  W, X, 
 \, 2\nabla_Y Z - [Y, Z]_C  +  \vev{Z,  \nabla_{Z_A} Y} Z^A \bigr) \\[1.0ex]
 = \ &   \Bigl\langle\, 2\nabla_W X - [W, X]_C \,,  
 \, 2\nabla_Y Z - [Y, Z]_C  +  \vev{Z,  \nabla_{Z_A} Y} Z^A
 \Bigr\rangle \\
 & 
 + \big\langle X , \nabla_{2\nabla_Y Z - [Y, Z]_C  +  \vev{Z,  \nabla_{Z_A} Y} Z^A}  W\big\rangle
 \\[1.0ex]
 = \ &  4\, \vev{\nabla_W X , \nabla_Y Z}    - 4 \vev{\nabla_W X, [Y, Z]_C}
 + \vev{[W, X]_C , [Y, Z]_C} \\
 & +  \vev{ Z , \nabla_{ 2\nabla_W X - [W, X]_C  } Y }  \\
 & + 2 \vev{ X , \nabla_{\nabla_Y Z}  W}  - \vev{X, \nabla_{[Y, Z]_C}W}
 + \vev{Z, \nabla_{Z_A} Y} \cdot \vev{X, \nabla_{Z^A} W } \;. 
\end{split}
\ee
Here we have used (\ref{AnablaB}) to obtain 
the term in the second line of the last equality. 
The same term, when expanded, gives terms that combine
with the first two of the last line.  We finally get (changing the overall
sign)
\be\label{secondbox}
\begin{split}
-\,{\cal T}(W,X,{\cal T}^{\uparrow}(Y,Z))\ =  \ & - 4 \vev{\nabla_W X , \nabla_Y Z}    + 4 \vev{\nabla_W X, [Y, Z]_C}
 - \vev{[W, X]_C , [Y, Z]_C}~~ \\
 & - 4 \vev{ Z , \nabla_{\nabla_W X}  Y}  +2 \vev{Z, \nabla_{[W, X]_C}Y}
 -  \vev{Z \nabla_{Z_A} Y} \, \vev{X \nabla_{Z^A} W }\;. 
\end{split}
\ee
Notice that the last two terms in here and in (\ref{curvred}) 
are the same and thus cancel out in the Bianchi identity.

Let us now compute the remaining term in (\ref{INVAlgBianchi99}), 
the covariant derivative of the torsion. Recalling the total 
antisymmetry that is left implicit and using (\ref{covderaction}) we get  
\be
\begin{split}
\nabla_W {\cal T} (X, Y, Z)  \ = \ & \ W \cdot {\cal T} (X, Y, Z) - 
3 {\cal T} (Y, Z, \nabla_W X)  \\
 = \ & \  W \cdot \vev{ 2\nabla_X Y - [X, Y]_C \,, Z}
 + W\cdot \vev{Y, \nabla_Z X}  \\
  & \ - 3 \Bigl(  \vev{2 \nabla_Y Z - [Y, Z]_C , \nabla_W X}
  + \vev{Z, \nabla_{\nabla_W X} Y } \Bigr) \;. 
\end{split}
\ee
Letting the $ W \cdot$ act inside the inner product for all terms except
the $\vev{[X, Y]_C , Z}$, using the metric compatibility (\ref{metriccomp}), 
and simplifying using the antisymmetry
one finds
\be
\begin{split}
\nabla_W {\cal T} (X, Y, Z)  \ 
 = \ & \  3\,\vev{\nabla_W\nabla_X Y , Z} \ - \ 3 \,\vev{\nabla_W X , \nabla_Y Z} 
-  W \cdot \vev{ [X, Y]_C \,, Z}  \\
  & \ + 3\, \vev{ \nabla_W X,  [Y, Z]_C}
  - 3\, \vev{\nabla_{\nabla_W X} Y , Z }  \;. 
\end{split}
\ee
Multiplying by $-4/3$ we have
\be\label{thirdbox}
\begin{split}
-{4\over 3}\, \nabla_W {\cal T} (X, Y, Z)  \ 
 = \ & \  -4\,\vev{\nabla_W\nabla_X Y , Z} \ + \ 4 \,\vev{\nabla_W X , \nabla_Y Z} 
+ {4\over 3} \,   W \cdot \vev{ [X, Y]_C \,, Z} ~ \\
  & \ -4 \, \vev{ \nabla_W X,  [Y, Z]_C}
  + 4\, \vev{\nabla_{\nabla_W X} Y , Z }\;.   
\end{split}
\ee
We can now add the three equations (\ref{curvred}), (\ref{secondbox}) and 
(\ref{thirdbox}) to find that
all terms except two cancel, 
\be
\label{wltfvm}
\begin{split}
{\cal R}(W,X,Y,Z)\, -\, {4\over 3}&\, (\nabla_{W}{\cal T})(X,Y,Z) \, - \, 
  {\cal T}(W,X,{\cal T}^{\uparrow}(Y,Z))   \ 
 \\
& \ =  \   {4\over 3} \,   W \cdot \vev{ [X, Y]_C \,, Z}  - \vev{[W, X]_C , [Y, Z]_C}\;. 
\end{split}
\ee
It remains to prove that the above right-hand side vanishes.  
For this we use (\ref{c-bracket-inner}), which gives
\be
\begin{split}
W \cdot \vev{X, [Y, Z]_C} \ =  \ & \   \vev{ [W, X]_C, [Y, Z]_C} 
+   \vev{ X , [W, [Y, Z]_C]_C}   \\
&  +   {1\over 2}\, [Y, Z]_C \cdot \vev{W, X} +  {1\over 2} X \cdot \vev{W, [Y, Z]_C} \;. 
\end{split}
\ee
Because of the implicit antisymmetry, the first term on the second
line vanishes 
and the second term on the second line can be moved
to the left-hand side,  
\be\label{STEp0234}
\begin{split}
{3\over 2} \, W \cdot \vev{X, [Y, Z]_C} \ =  \ & \   \vev{ [W, X]_C, [Y, Z]_C} 
+   \vev{ W , [[X, Y]_C, Z]_C}\;,  \end{split}
\ee
where we also used the total antisymmetry in the last term. 
This last term is the C-Jacobiator (\ref{c-jacobiator}) which, 
assuming multiplication by  antisymmetric projectors, reads  
\be
J_C (X, Y, Z) \ \equiv \ 3 [[X, Y]_C , Z]_C \;, 
\ee
so that (\ref{STEp0234}) becomes 
\be
\label{kavm}
\begin{split}
{3\over 2} \, W \cdot \vev{[X, Y]_C,   Z} \ =  \ & \   \vev{ [W, X]_C, [Y, Z]_C} 
+ {1\over 3}   \vev{ W ,J_C (X, Y, Z) } \;. 
\end{split}
\ee
Using (\ref{c-jacobiator}) and the antisymmetry we can
use
\be
J_C (X, Y, Z) \ = \   {1\over 2} \,\vec{\partial}\, \vev{  [X, Y]_C ,  Z}
\quad \Rightarrow \quad
\vev{W, J_C (X, Y, Z)} \ = \   {1\over 2} \, W \cdot \vev{ [X, Y]_C ,  Z}\;. 
\ee
Inserting this in (\ref{kavm}) we get  
\be
\label{kasvm}
\begin{split}
{3\over 2} \, W \cdot \vev{[X, Y]_C \,,  Z} \ =  \ & \   \vev{ [W, X]_C, [Y, Z]_C} 
+ {1\over 6}  \,W\cdot  \vev{ [X, Y]_C ,  Z}\;, 
\end{split}
\ee
so we finally have 
\be
\label{kassvm}
{4\over 3} \, W \cdot \vev{[X, Y]_C \,,  Z} \ =  \   \vev{ [W, X]_C, [Y, Z]_C} \;. 
\ee
This is the desired identity, and therefore the right-hand side
of (\ref{wltfvm}) vanishes and we have proven the algebraic
Bianchi identity.

\section{Connection with generalized geometry}
In this section we make contact with results in the 
literature on generalized geometry. This introduces the generalized 
metric. It will also be useful below when we compare with the frame formalism of Siegel.

\subsection{Generalized metric}\label{gengeometry1}
We now introduce the generalized metric. In this subsection we closely follow the treatment in 
generalized geometry as given by Gualtieri in \cite{Gualtieri:2007bq}. 
We first note that on the (generalized) 
tangent space $TM$ with $O(D,D)$ metric $\langle \cdot \,, \cdot \rangle$ we can 
select a $D$-dimensional basis $C_+$ which is 
positive definite with respect to the inner product
 $\langle \cdot \,, \cdot \rangle$.  This choice corresponds to a choice of generalized metric. The orthogonal complement $C_-$
 is also $D$-dimensional and negative definite.\footnote{In double field theory we often 
 consider the full spacetime metric, i.e., a metric of Lorentzian signature. Then we should consider 
 a decomposition into subspaces $C^{\pm}$ each of Lorentzian signature (but with opposite overall sign). This generalization 
 is straightforward and so we spell out only the details for positive definite signature.} 
 We thus have 
 \be\label{splitting}
 TM \ = \ C_+ \oplus C_-\,.
 \ee
For arbitrary vectors $X, Y\in TM$ we write decompositions
\be
X \ = \ X_+ + X_- \,, ~~~ Y \ = \ Y_+ + Y_- \,.
\ee
We now define the generalized metric tensor ${\cal H}(X,Y)$ by the relation  
\be
{\cal H} (X, Y) \ \equiv \ \vev{X_+, Y_+}  \ - \ \vev{X_- , Y_-} \,. 
\ee
This is sometimes written schematically as
\be
{\cal H} \ = \  \langle \ , \ \rangle \bigl\vert_{C_+} 
- \  \langle \ , \ \rangle \bigl\vert_{C_-} \;. 
\ee
We can use the orthogonality of the subspaces $C_+$ and $C_-$ to write
\be
\begin{split}
{\cal H} (X, Y) \ = \ & \ \vev{X_+ + X_-, Y_+}  \ - \ \vev{X_+ 
+ X_- , Y_-} \\
\ = \ & \ \vev{X, Y_+}  \ - \ \vev{X , Y_-} \;, 
\end{split}
\ee
and we conclude that 
\be
\label{vmvmvg}
{\cal H} (X, Y) \ =   \ \vev{X, Y_+- Y_-}   \ = \ \vev
{ X_+ - X_-, Y}\;.   \ee
We now define the linear operator $S$ by
\be
S X  \ \equiv \  X_+ - X_- \,,
\ee
i.e., it changes the overall sign of the part in $C_{-}$ but leaves $C_+$ invariant. 
It follows that the operator squares to one and that it preserves
the $C_\pm$ spaces:
\be\label{Ssquare}
S^2 \ = \ {\bf 1} \,, ~~~~  S X_{\pm } \ = \ \pm X_{\pm} \,.
\ee
This allows us to rewrite (\ref{vmvmvg}) as
\be
\label{vmswvg}
{\cal H} (X, Y) \ =   \ \vev{X, SY}   \ = \ \vev
{ SX, Y} \;,              
\ee
demonstrating that $S$ is a symmetric map.  It is also an automorphism since by (\ref{Ssquare}) 
\be
\vev{ SX \,, \; SY }  \ = \ \vev{X\,, S^2Y} \ = \ \vev{X, Y} \;. 
\ee

\medskip
Let us now phrase the covariant constancy of the generalized metric in these invariant terms.  
We want to impose 
\be
\nabla {\cal H}  \ = \ 0 \,,  ~~~\hbox{and} ~~~ \nabla {\cal \eta} \ = \ 0 \,,
\ee
where the last condition is equivalent to (\ref{metriccomp}). 
We compute for the first with (\ref{covderaction}) 
\be
\label{vmswtvg}
\begin{split}
0\ = \ \nabla {\cal H} (X, Y, Z) \ = \ &  \nabla_Z {\cal H} (X, Y) \\
= \ &  Z \cdot  {\cal H} (X, Y)  -  {\cal H} (\nabla_ZX, Y) 
-  {\cal H} (X, \nabla_ZY) \\
=  \ &  Z \cdot  \vev{X, SY}  -  \vev{\nabla_ZX, SY} 
-  \vev{X, S\nabla_ZY}\;. 
\end{split}
\ee
The relation $\nabla{\cal \eta} = 0 $ implies with (\ref{metriccomp}) 
\be
Z \cdot  \vev{X, SY} \ = \ \vev{\nabla_Z X, SY} + \vev{X, \nabla_Z SY}\;. 
\ee
Insertion into (\ref{vmswtvg}) then yields 
\be
0 \ = \ \vev{X, \nabla_Z SY} \ - \ \vev{X, S\nabla_Z Y}\;. 
\ee
We thus learn that the linear map $S$ commutes with
covariant differentiation, 
\be
\label{commvm}
 \nabla_X  S \ = \ S\, \nabla_X \,.
\ee
Thus the spaces $C_+$ and $C_-$ are preserved by
covariant differentiation:
\be
\label{presvm}
\nabla_X:  \; C_\pm \to C_\pm \,, ~~~\forall \, X \,. 
\ee
It is also simple to see that  for certain inputs the C- and D-brackets coincide:
\be
\label{ceqd}
[X_+ , Y_-]_D  \ = \ [X_+, Y_-]_C \,.
\ee
This follows because the  
C- and D- brackets differ by a term proportional to the derivative of 
$ \vev{X_+, Y_-} = 0$,  
as we can see from (\ref{covDbracket}).

\subsection{Implications for connections and curvature}\label{connectionimpl}
We derive now some conclusions for the connections and our  curvature tensor
with respect to the splitting $TM =C_+ \oplus C_-$.  
 Let us first summarize the constraints 
imposed so far and introduce a final one, item (3) below, that introduces the dilaton:
 \begin{itemize}
  \item[(1)] The generalized torsion vanishes, ${\cal T}(X,Y,Z) \ = \ 0$ for all $X,Y,Z$.
  \item[(2)] The $O(D,D)$ metric and the generalized metric are covariantly constant,   
   \be
    \nabla {\cal H}  \ = \ 0 \,, \qquad  \nabla {\cal \eta} \ = \ 0 \,.
  \ee
  \item[(3)]   
The density $e^{-2d}$  
   allows for integration by parts as 
    \be
     \int e^{-2d}\,f\,{\rm div}V \ = \ -\int e^{-2d}\,V f\;, 
   \ee
for any function (scalar) $f$ and vector $V$, where the divergence is defined as 
 \be
  {\rm div}{V} \ = \ \vev{\nabla_{Z_{A}}V,Z^{A}} \;.
 \ee
 \end{itemize}  

Next we derive some useful relations for various connection components, 
written using the splitting of the tangent bundle. 
We first note that the torsion constraint allows some 
nice simplification. With (\ref{defgentor}) we infer  
\be
\label{defgentor99}
0\ = \  {\cal T} (X_+, Y_-, Z) \ 
\equiv  \    \vev{ \nabla_{X_+} Y_- - \nabla_{Y_-} X_+- [X_+, Y_-]_D\, , \,Z } + \vev{Y_-, \nabla_Z X_+}  \,. 
\ee
The last term vanishes and the above holds for all $Z$,  so that
we find
\be
\label{torsym}
\nabla_{X_+} Y_-  \ - \ \nabla_{Y_-} X_+  \ = \ [ X_+ \,, Y_-]_D \ = \ 
[ X_+ \,, Y_-]_C \, ~.
\ee
In order to gain further insights  we start from the metric compatibility 
\be
\label{dfldfk}
X\cdot  \vev{Y, Z} = \vev{\nabla_XY, Z}  + \vev{Y, \nabla_XZ}\;, 
\ee
so that using the torsion constraint for the second term  we find
 \be
X\cdot  \vev{Y, Z} = \vev{\nabla_XY, Z}  + \vev{Y, \nabla_Z X}
+ \vev{Y, [X, Z]_D } - \vev{ Z, \nabla_Y X} \;. 
\ee
Moving the third term on the right-hand side to the left, and
using (\ref{derdorf}), we get 
 \be
 \label{vmans}
\vev{[X, Y]_D, Z } = \vev{\nabla_XY, Z}  + \vev{Y, \nabla_Z X}
 - \vev{ Z, \nabla_Y X} \;. 
\ee
Making choices for which only the first term in the above
right-hand side is nonzero we find
\be
\label{concomp}
\begin{split}
\vev{\nabla_{X_-} Y_+ \,, Z_+ } \ = \ & \vev{ \,[X_-, Y_+]_D, Z_+}\;,  \\
\vev{\nabla_{X_+} Y_-\,, Z_- } \ = \ & \vev{ [X_+, Y_-]_D, Z_-} \;. 
\end{split}
\ee
These equations explicitly determine the corresponding projections 
of the connection in terms of the D-bracket.
Since $\nabla_{X_-} Y_+$ does not 
have a component in $C_-$ and $\nabla_{X_+} Y_-$
does not have a component in $C_+$ (see (\ref{presvm})),
back to (\ref{concomp}) we see that
\be
\label{-+-conn}
\begin{split}
\nabla_{X_-} Y_+\ = \ \ & [X_-, Y_+]_{D+} \,,\\ 
\nabla_{X_+} Y_-\,\ = \ \ & [X_+, Y_-]_{D-} \,. ~
\end{split}
\ee
We could also trade the above D-brackets for C-brackets.
As a consistency check we can also confirm that the torsion constraint (\ref{torsym}) 
is  satisfied.  Indeed, using the above expressions and recalling
that for $C_+, C_-$ inputs the D-bracket is antisymmetric we find
\be\label{offcovcurl}
\begin{split}
\nabla_{X_+} Y_-  \ - \ \nabla_{Y_-} X_+  \ &= \ 
[X_+, Y_-]_{D-}  -  [Y_-, X_+]_{D+} \\
\ &= \ 
[X_+, Y_-]_{D-}  +  [X_+, Y_-]_{D+} \\
\ &= \ [X_+, Y_-]_D\,. 
\end{split}
\ee

So far we have derived relations that determine certain projections of the connection 
that are `off-diagonal' with respect to the decomposition $C_+\oplus C_-$. 
Next, we derive an equation that determines 
 the totally antisymmetric part 
of the connections.  
We begin with (\ref{vmans}) and rewrite the last
term using metric compatibility, 
 \be
 \label{vmansdl}
 \begin{split}
\vev{[X, Y]_D, Z } \ = \ &\vev{\nabla_XY, Z}  + \vev{Y, \nabla_Z X}
 - \vev{ Z, \nabla_Y X} \\
 \ = \ &\vev{\nabla_XY, Z}  + \vev{ \nabla_Z X, Y}
 - Y\,  \vev{ Z, X}  + \vev{\nabla_Y Z, X}  \;. 
 \end{split}
 \ee
It follows that 
 \be 
 \vev{[X, Y]_D, Z }   \ +  \  Y\, \vev{ Z, X}\ = \  \sum_{X, Y, Z}^{cyc}
 \vev{\nabla_XY, Z}\;. 
\ee
The left-hand side is not manifestly cyclic but it is cyclic. Indeed, 
\be
\begin{split}
\vev{[X, Y]_D, Z } \ +  \ Y\, \vev{ Z, X}\ = \  &\ 
\vev{[X, Y]_D, Z } \ +  \ \vev{[Y, Z]_D, X} \ + \  
\vev{Z, [Y, X]_D }  \\
\ = \  &\ \vev{[Y, Z]_D, X} \ + \ 
\vev{\,[X, Y]_D +  [Y, X]_D \, , Z }   \\
\ = \  &\ 
 \vev{[Y, Z]_D, X} \ + \ Z\, \vev{X, Y}  \,,
\end{split}
\ee
where we used (\ref{indfreed}) in the second line in order to 
rewrite the symmetric part of the D-bracket. This proves 
the cyclicity in $X,Y,Z$. We therefore have
\be
\label{totant}
 \sum_{X, Y, Z}^{cyc}  \vev{\nabla_XY, Z} \ = \ 
 {1\over 3}   \sum_{X, Y, Z}^{cyc}  
 \Bigl( \vev{[X, Y]_D, Z } \ +  \ X\, \vev{ Y, Z} \, \Bigr)\;, 
\ee
which determines the totally antisymmetric (cyclic) part 
of the connection.

Let us summarize and interpret our above results.  
First, for the connection coefficients $\Gamma$ in a coordinate basis, defined in (\ref{Christoffel-like}), 
the relation (\ref{totant}) shows that the totally antisymmetric 
part  vanishes, $\Gamma_{[MNK]}=0$, because on the right-hand side the
$O(D,D)$ metric is constant and the D-bracket is zero. 
Second, off-diagonal projections of the connections are determined by (\ref{-+-conn}). This leaves a 
`Hook-like' Young tableau in the connections coefficients undetermined, but whose trace part is 
determined by the dilaton according to constraint (3) above. This leaves the traceless part of this 
representation undetermined; the connections cannot be determined completely by means of covariant 
constraints. 
 
It is instructive to write and count the connection components with respect to a coordinate basis, 
also to make contact with the explicit results in   \cite{Hohm:2011si}. 
First we have to introduce some notation. Because of $S^2={\bf 1}$, see (\ref{Ssquare}), one can define projection 
operators $P_{\pm}$, mapping $P_{\pm}(TM)=C_{\pm}$, by 
 \be
  P_{\pm} \ = \ \frac{1}{2}\left({\bf 1}\pm S\right)\;, 
 \ee
so that $P_{\pm}^2=P_{\pm}$ and $P_+ P_- =0$.   For any $O(D,D)$ tensor $V$ we introduced  in 
\cite{Hohm:2011si} the notation\footnote{In order to compare with \cite{Hohm:2011si} set $P_+\equiv \bar{P}$ and $P_-\equiv P$.} 
 \be
  V_{\bar{M}} \ \equiv \ (P_+)_{M}{}^{N} V_{N}\;, \qquad V_{\,\nin{M}} \ \equiv \ (P_-)_{M}{}^{N} V_{N}\;, 
  \quad {\rm etc.}
 \ee   
Contracting now the defining relation (\ref{Christoffel-like}) for the Christoffel symbols with $P_+$ and 
$P_-$ and 
employing this notation we obtain 
 \be
  (P_+)_M{}^{K}(P_-)_N{}^{P}\nabla_{\partial_K}(\partial_P) \ = \ \Gamma_{\bar{M}\,\nin{N}}{}^{P}\partial_P\;.
 \ee 
Moving the projectors inside the covariant derivatives, remembering (\ref{scaling}), yields 
 \be\label{TORstep1}
  \nabla_{(P_+ \partial)_M}(P_- \partial)_N-(P_+)_M{}^{K}\partial_K (P_-)_N{}^{P}\partial_P
  \ = \ \Gamma_{\bar{M}\,\nin{N}}{}^{P}\partial_P\;, 
 \ee 
where $(P_+ \partial)_M =
(P_+)_{M}{}^{N}\partial_N$, etc. Completely analogously we have 
 \be\label{TORstep2}
  \nabla_{(P_- \partial)_N}(P_+ \partial)_M-(P_-)_N{}^{K}\partial_K (P_+)_M{}^{P}\partial_P
  \ = \ \Gamma_{\,\nin{N}\bar{M}}{}^{P}\partial_P\;. 
 \ee 
Subtracting (\ref{TORstep2}) from (\ref{TORstep1}) we obtain 
 \be\label{NABlaCurl}  
 \begin{split}
  \nabla_{(P_+ \partial)_M}(P_- \partial)_N - \nabla_{(P_- \partial)_N}(P_+ \partial)_M
  \ = \ 
  & \bigl[  (P_+)_M{}^{K}\partial_K (P_-)_N{}^{P}
  -(P_-)_N{}^{K}\partial_K (P_+)_M{}^{P}\bigr] \partial_P \\
  &+\big(\Gamma_{\bar{M}\,\nin{N}}{}^{P}- \Gamma_{\,\nin{N}\bar{M}}{}^{P}\big)\partial_P\;.  
 \end{split}
 \ee 
On the other hand, by (\ref{offcovcurl}) the left-hand side can also be written with the D-bracket, 
 \be\label{DEXpression}
  \nabla_{(P_+ \partial)_M}(P_- \partial)_N - \nabla_{(P_- \partial)_N}(P_+ \partial)_M
  \ = \ \big[P_{+M},P_{-N}\big]_{D}\;, 
 \ee 
where we interpret $P_{+M}{}^{N}$ as a generalized vector with vector index $N$, treating $M$
as a pure label index. Using (\ref{dorfbrakcc}) this D-bracket reads   
 \be  
  \big[P_{+M},P_{-N}\big]_{D} \ = \ \big((P_+)_M{}^{K}\partial_K (P_-)_{N}{}^{P}
  - (P_-)_{N}{}^{K} \partial_K(P_+)_{M}{}^{P} 
  +\partial^P(P_+)_{MK}(P_-)_{N}{}^{K}\big)\partial_P\;.
 \ee 
This has to be equal to the right-hand side of (\ref{NABlaCurl}) and so we have arrived at a 
relation determining $\Gamma$, 
 \be
  \Gamma_{\bar{M}\,\nin{N}P}- \Gamma_{\,\nin{N}\bar{M}P} \ = \   
  \partial_P(P_+)_{MK} \,(P_-)_{N}{}^{K}\;, 
 \ee
where we dropped the basis vectors $\partial_P$ and lowered the index $P$. We see that various terms cancelled.   
Since the totally antisymmetric part of $\Gamma$ vanishes we can rewrite the left-hand side as
 $\Gamma_{\bar{M}\,\nin{N}P}+ \Gamma_{\,\nin{N}P\bar{M}}=-\Gamma_{P\bar{M}\,\nin{N}}$, so that we finally get 
  \be\label{metriconncomp}
   \Gamma_{P\bar{M}\,\nin{N}} \ = \ -(P_-)_{N}{}^{K} \partial_P(P_+)_{MK} \;,  
  \ee 
which is in agreement with eq.~(2.54) in \cite{Hohm:2011si}.  

We close the discussion of the connection components by counting the number of 
undetermined connections. Without any constraints, 
$\Gamma_{MNK}$ has $(2D)^3 =8D^3$ 
components. 
We next subtract the numbers of independent constraints, which will give the number of 
undetermined connections:
 \begin{itemize}
  \item Metric compatibility:   
   \be
    \Gamma_{M(NK)} \ = \ 0\; : \qquad 2D\cdot  \frac{2D(2D+1)}{2} \ = \ \frac{1}{2} \cdot 8 D^3+2D^2\;. 
   \ee
  \item Vanishing torsion: Using metric compatibility, the torsion components ${\cal T}_{MNK}$
  are totally antisymmetric. Thus, the number of constraints is 
   \be
    \frac{2D(2D-1)(2D-2)}{6} \ = \ \frac{1}{6} \cdot 8D^3-2D^2+\frac{2}{3}D\;. 
   \ee  
  \item  Covariant constancy of ${\cal H}$: the independent components determined by this constraint 
  are given by  (\ref{metriconncomp}), so that we obtain the number  $(2D) D^2=2D^3$. 
   \item 
    Trace constraint and dilaton: this constraint determines the trace part $\Gamma_{KN}{}^{K}$, 
    thus adding $2D$ constraints.  
 \end{itemize}  
In total, the number of undetermined connections is given by 
 \be
  8D^3- \Big(\frac{1}{2} 8 D^3+2D^2\Big)-\Big(\frac{1}{6} 8D^3-2D^2+\frac{2}{3}D\Big) -2D^3-2D
  \ = \ \frac{2}{3}D(D^2-4)\;, 
 \ee
in agreement with the counting in  \cite{Hohm:2010xe} and in agreement with the dimension
of two  traceless Hook tableau representations of $GL(D)$.

\medskip

Let us finally derive some conclusions for certain projections of 
the generalized Riemann tensor. This tensor ${\cal R}(X,Y,Z,W)$ defined in (\ref{curvaturehere_it_is}) 
is antisymmetric under the exchange
of $X$ and $Y$ as well as under the exchange of $Z$ and $W$.
We can now give an alternative formula using the Courant bracket 
by making the symmetries explicit by the relation
\be
{\cal R} (X, Y, Z, W) \ = \ {1\over 4} \Bigl( 
{\cal R} (X, Y, Z, W) - {\cal R} (Y, X, Z, W) - {\cal R} (X, Y, W, Z)
+ {\cal R} (Y, X, W, Z) \Bigr) \;. 
\ee
Since the D-bracket differs from the C-bracket only by a term symmetric 
in the arguments we can replace in  
the right-hand side the D-bracket by the C-bracket. One finds
\be
\label{curvaturehere_it_is_vm_cour} 
\begin{split}
{\cal R} (X, Y, Z, W)  \ \equiv & \ \ \ ~ ~\vev{(\nabla_X \nabla_Y -  \nabla_Y \nabla_X- \nabla_{[X, Y]_C} ) Z \,, W} \\
& \ \  + \vev{(\nabla_Z \nabla_W -  \nabla_W \nabla_Z - \nabla_{[Z, W]_C}) X \,, Y}  ~~\\
& \ \  +  \frac{1}{4}\Bigl( \,
\vev{Y, \nabla_{Z_A} X}  \, \vev{W, \nabla_{Z^A} Z }
\ -\  \vev{X, \nabla_{Z_A} Y}  \, \vev{W, \nabla_{Z^A} Z } 
\\[-1.8ex] 
& \ \  \hskip30pt -
\vev{Y, \nabla_{Z_A} X}  \, \vev{Z, \nabla_{Z^A} W }
\ +\  \vev{X, \nabla_{Z_A} Y}  \, \vev{Z, \nabla_{Z^A} W }\Bigr)\;,   \\ 
\end{split}
\ee
where we used (\ref{adqui}), that holds for the C- and the D-bracket. 
Note that the equivalence of the two expressions for the curvature ${\cal R}$ 
only holds when the connection is compatible with the $O(D, D)$ metric.
We now see that 
\be
\label{vmlck}
{\cal R} (X_+, Y_-, Z, W)  \ = \ \vev{(\nabla_{X_+} \nabla_{Y_-} -  \nabla_{Y_-} \nabla_{X_+} - \nabla_{[X_+, Y_-]_C} ) Z \,, W}\,, 
\ee
since the covariant derivatives preserve the orthogonal $+$ and $-$ subspaces 
and so only the first line in (\ref{curvaturehere_it_is_vm_cour}) is non-zero. 
As a simple consequence
\be\label{simpcons}
{\cal R} (X_+, Y_-, Z_+, W_-) \ = \ 0 \,.
\ee
Using the algebraic Bianchi identity we also see immediately that
\be
{\cal R} (X_+, Y_+, Z_-, W_-) \ = \ 0 \,.
\ee

\section{Relation to frame formalism}
In this section we evaluate the geometrical quantities with respect to a frame basis 
in order to make contact with the frame formalism of Siegel. In particular, we will show 
how the constraints of Siegel are recovered from our constraints above and that the 
Riemann tensor reduces to the curvature of the frame formalism.

\subsection{Generalities}
We introduce a general (`frame') basis $E_{A} = E_{A}{}^{M}\partial_{M}$, with $A = 1, 2, \ldots, 2D$
and with $E_M{}^A$, defined to be the inverse of $E_A{}^M$, assumed to exist.  
The frame basis  
is not necessarily 
orthonormal, so we define 
 \be
  {\cal G}_{AB} \ \equiv \ \vev{E_{A},E_{B}}\;.
 \ee
We will assume that the basis $E_A$ 
 respects the decomposition (\ref{splitting}) 
of the tangent space into $C_+ \oplus C_-$. More explicitly,
the basis decomposes
as $E_{A}=(E_{a},E_{\bar{a}})$, where indices $a,b=1,\ldots, D$ refer to $C_-$ and indices 
$\bar{a},\bar{b}=1,\ldots,D$ refer to $C_+$,\footnote{The convention for 
unbarred and barred indices here is  chosen such that it complies with that of \cite{Hohm:2010xe}.} 
so that for the $O(D,D)$ invariant inner product 
we have the constraints   
 \be\label{calG}
 {\cal G}_{a\bar b} \ = \  
   \vev{E_{a},E_{\bar{b}}} \ = \ 0\;, \qquad \det ({\cal G}_{ab})<0\;, \quad \det( {\cal G}_{\bar{a}\bar{b}})>0\;. 
 \ee
 We also define  
 \be
 \label{Hab}
 {\cal H}_{AB}  \ \equiv \ {\cal H} (E_A, E_B) \,.
 \ee
With respect to the  basis $E_A$  
 we then introduce spin 
connection components $\omega_{AB}{}^C$  
as 
 \be\label{spincomp}
  \nabla_{E_{A}}E_{B} \ = \ -\omega_{AB}{}^{C}E_{C}\;, 
 \ee
similar to the Christoffel-like connections (\ref{Christoffel-like}). 
In physicists terminology the spin connections are related to the Christoffel connections 
via a `vielbein postulate' that states that the `vielbein' or frame components $E_{A}{}^{M}$ 
are covariantly constant with respect to the simultaneous use of the Christoffel and spin connection. 
In this invariant formulation this is not an independent postulate but rather follows directly: 
 \be
 \begin{split}
  \nabla_{E_{A}}E_{B} \ &= \  E_{A}{}^{M}\nabla_{\partial_M}\big(E_{B}{}^{N}\partial_{N}\big) \ = \ 
  E_{A}{}^{M}\big(\partial_{M}E_{B}{}^{N}\,\partial_{N}+E_{B}{}^{N}\nabla_{\partial_{M}}(\partial_{N})\big) \\
  \ &= \ E_{A}{}^{M}\big(\partial_{M}E_{B}{}^{N}+\Gamma_{MK}{}^{N}E_{B}{}^{K}\big)\partial_{N} \ \equiv \ 
  -\omega_{AB}{}^{C}E_{C}{}^{N}\partial_{N}\;, 
 \end{split} 
 \ee
where we used (\ref{Christoffel-like}) and (\ref{spincomp}). Bringing the right-hand side to the left-hand side, we get 
 \be\label{vielbeinpost}
    E_{A}{}^{M}\big(\partial_{M}E_{B}{}^{N}+\Gamma_{MK}{}^{N}E_{B}{}^{K}+\omega_{MB}{}^{C}E_{C}{}^{N}\big)\partial_{N}    
    \ = \ 0 \;. 
 \ee 
This implies  the vielbein postulate in the usual form       
 \be\label{VielBeinPost}
  \nabla_{M}E_{A}{}^{N} \ \equiv \ \partial_M E_{A}{}^{N}+\Gamma_{MK}{}^{N}E_{A}{}^{K}+\omega_{MA}{}^{B}E_{B}{}^{N} \ = \ 0\;. 
 \ee
Here we introduced the notation $\nabla_M$ for the covariant derivative with respect to both the Christoffel 
and spin connection, which acts in the usual way on tensors with an arbitrary number of curved and flat indices. 
We will also write $\nabla_{A}=E_{A}{}^{M}\nabla_{M}$.  
Let us stress that here and below we employ the physicists notation for covariant derivatives with a pure letter as an index, 
as opposed to covariant derivatives like $\nabla_{\partial_M}$ used in the nomenclature of mathematicians.  

Finally, let us discuss the generalized metric in this basis. 
As $S(X_{\pm})=\pm X_{\pm}$ for the endomorphism introduced in sec.~\ref{gengeometry1}
we have 
 \be
  S(E_{a}) \ = \ -E_{a}\;, \qquad S(E_{\bar{a}}) \ = \ E_{\bar{a}}\;.
 \ee
Therefore, the frame components of the generalized metric as defined in   
(\ref{vmswvg}) are given by 
 \be\label{flatcomp}
 \begin{split}
 {\cal H}_{ab} \ = \ 
  {\cal H}(E_a,E_b) \ &= \ \vev{ E_{a},S(E_{b})} \ = \ -\vev{ E_{a},E_{b}} \ = \ -{\cal G}_{ab}\;, \\
 {\cal H}_{\bar a \bar b} \ = \  {\cal H}(E_{\bar a},E_{\bar b}) \ &= \ \vev{ E_{\bar a},S(E_{\bar b})} \ = \  \vev{ E_{\bar{a}},E_{\bar{b}}} \ = \ {\cal G}_{\bar{a}\bar{b}}\;. 
 \end{split}
 \ee
Moreover, since $S$ preserves the orthogonal subspaces we have 
  \be
  {\cal H}_{a \bar b} \ = \ 
    {\cal H}(E_a,E_{\bar b}) \ = \ 0\;.
  \ee
We can finally express the generalized metric in a coordinate basis in terms of the frame components. We have 
 \be
  {\cal H}_{MN} \ \equiv \ {\cal H}(\partial_M,\partial_N) \ = \ {\cal H}(E_{M}{}^{A}E_{A}\,,\, E_{N}{}^{B}E_B) \ = \ 
  E_{M}{}^{A} E_{N}{}^{B} {\cal H}_{AB}\;,
 \ee
where $E_{M}{}^{A}$ denotes the inverse of $E_{A}{}^{M}$. 
Inserting the non-vanishing components (\ref{flatcomp}) 
of ${\cal H}_{AB}$ we get 
 \be
  {\cal H}_{MN}   
   \ = \ E_{M}{}^{\bar{a}} E_{N}{}^{\bar{b}} {\cal G}_{\bar a \bar b}  -E_{M}{}^{a} E_N{}^b
{\cal G}_{ab}    \ = \ E_{M}{}^{\bar{a}} E_{N\bar{a}} -E_{M}{}^{a} E_{Na}\;,
 \ee
where indices are contracted with ${\cal G}_{AB}$.  This coincides with the expressions for the generalized 
metric in terms of the frame fields given in \cite{Hohm:2010pp,Hohm:2010xe}.

\subsection{Constraints}
We now give the various constraints in the frame formalism 
and show that they are equivalent to those in Siegel's 
original approach. 
We start with the constraint stating compatibility 
of the connection with the generalized metric. According to (\ref{presvm}) this constraint is equivalent to 
the condition that the connection preserves the subspaces $C_{\pm}$. In the frame basis $(E_{a},E_{\bar{a}})$
this amounts to the constraint that the connection coefficients (\ref{spincomp}) are only non-vanishing 
for $\omega_{Aa}{}^{b}$ and $\omega_{A\bar{a}}{}^{\bar{b}}$, and that there are no `off-diagonal' components.  
Put differently, the structure group takes the factorized form $GL(D)\times GL(D)$. In Siegel's formalism 
this is assumed from the outset, so in this formulation covariant constancy of the generalized metric is automatic.  

Next, we inspect the constraint (\ref{metricity}) stating compatibility of the $O(D,D)$ metric $\vev{\, ,\, }$ and the connection, 
 \be\label{metricity}
\nabla_Z\, \vev{X, Y}  \ = \  Z\,\vev{X, Y}  \ = \ \vev{ \nabla_Z X , Y}  + \vev{X, \nabla_Z Y}\;.
\ee
Specialized to the 
anholonomic basis $E_{A}$ it implies
 \be
 \begin{split}
  E_{A}\, \vev{E_{B},E_{C}} \ = \ E_{A}{\cal G}_{BC} \ &= \ \vev{ \nabla_{E_{A}} E_{B} , E_{C}}  + \vev{E_{B}, \nabla_{E_{A}} E_{C}} \\
  \ &= \ -\omega_{AB}{}^D {\cal G}_{DC} 
  -\omega_{AC}{}^D{\cal G}_{DB}\;, 
 \end{split}
 \ee 
and thus   
 \be\label{nablaG}
 \begin{split}  
  \nabla_{A}{\cal G}_{BC} \ \equiv &\  \ E_{A}{\cal G}_{BC}+\omega_{AB}{}^D {\cal G}_{DC} 
  +\omega_{AC}{}^D{\cal G}_{DB}   \\ 
   = & \ \  E_{A}{\cal G}_{BC}+\omega_{ABC}
  +\omega_{ACB}  \ = \ 0 \,.
  \end{split} 
 \ee
This is the covariant constancy of the tangent space metric imposed in Siegel's frame formalism as one 
of the constraints. 

Finally, we inspect the constraint of vanishing torsion.  The torsion tensor (\ref{defgentor})  evaluated for the basis $E_{A}$ reads 
 \be\label{TorsionStep1}
   {\cal T}(E_{A},E_{B},E_{C}) \ = \ \vev{\nabla_{E_{A}}E_{B}- \nabla_{E_{B}}E_{A}-\big[E_{A},E_{B}\big]_{D},E_{C}}
  +\vev{E_{B},\nabla_{E_{C}}E_{A}} \;.
 \ee 
In order to compare this with the torsion constraint in the frame formalism we introduce some notation.
As in \cite{Hohm:2010xe} we define 
generalized `coefficients of anholonomy'  
$\Omega_{AB}{}^{C}$ by use of the C-bracket:  
 \be\label{anholonomy}
  \big[E_{A},E_{B}\big]_{C} \ = \ \Omega_{AB}{}^{C} E_{C}\;, 
 \ee
where we stress again that this equation holds generally, not only when acting on functions 
satisfying the strong constraint. With (\ref{covDbracket}) we then find for the D-bracket 
 \be\label{Dbracket}
  \big[E_{A},E_{B}\big]_{D} \ = \ \big[E_{A},E_{B}\big]_{C} 
  +\frac{1}{2}(E_{C}\,\vev{E_A,E_B}) \,E^C  
  \ = \ \big( \Omega_{AB}{}^{C}+\frac{1}{2} E^{C}{\cal G}_{AB}\big)E_{C}\;.  
 \ee  
Inserting this into (\ref{TorsionStep1}) and using (\ref{spincomp}) we find for the torsion 
 \be
 \begin{split}
   {\cal T}(E_{A},E_{B},E_{C}) \ &= \ -\omega_{ABC}+\omega_{BAC}-\Omega_{ABC}-\frac{1}{2}E_{C}{\cal G}_{AB}-\omega_{CAB} \\
  \ &= \ -\big(\Omega_{ABC}+2\big(\omega_{[AB]C}+\frac{1}{2}\omega_{C[AB]}\big)\big)-\frac{1}{2}\big(E_{C}{\cal G}_{AB}
  +2\omega_{C(AB)}\big)\;.  
 \end{split} 
 \ee
Comparing with eq.~(2.21) of  \cite{Hohm:2010xe} we infer 
 \be
  {\cal T}(E_{A},E_{B},E_{C}) \ = \ -\, {\cal T}_{ABC}^{\rm S}-\frac{1}{2}\nabla_{C}{\cal G}_{AB}\;, 
 \ee
where ${\cal T}^{\rm S}$ denotes the torsion tensor of Siegel. Since we assume the 
metricity constraint (\ref{nablaG}) it follows that vanishing generalized  
 torsion is equivalent 
to the zero  
 torsion constraint in the frame formalism. 
Moreover,  constraint (3) in sec.~\ref{connectionimpl}, determining the trace of the connection 
in terms of the dilaton, coincides with one of the constraints in Siegel's frame formalism. 
Summarizing, all constraints imposed 
here agree with the constraints in the frame formalism.

We close this subsection by giving the explicit spin connection components solving the above 
constraints, which can be determined immediately from the results in sec.~\ref{connectionimpl}. 
First,  specializing the first equation in (\ref{-+-conn}) to the frame basis we find 
 \be  
  \nabla_{E_a}E_{\bar b} \ \equiv \ -\omega_{{a}\bar b}{}^{\bar c}E_{\bar c} 
  \ = \ \big[E_{{a}},E_{\bar b}\big]_{C+} \ = \ \Omega_{a\bar b}{}^{\bar c}E_{\bar c}\;,
 \ee
where we noted that for  off-diagonal projections we can replace the D-bracket by the C-bracket, 
and we inserted (\ref{anholonomy}). From this we conclude 
that 
$\omega_{{a}\bar b}{}^{\bar c}=-\Omega_{{a}\bar b}{}^{\bar c}$.
The analogous relation for the opposite projection follows from the second  equation in (\ref{-+-conn}), and so we have 
in total 
 \be\label{offconnec}  
  \omega_{a\bar{b}}{}^{\bar{c}} \ = \ -\Omega_{a\bar{b}}{}^{\bar{c}}\;, 
   \qquad \omega_{\bar{a}b}{}^{c} \ = \ -\Omega_{\bar{a}b}{}^{c}\;,
 \ee
which agrees with eq.~(2.34) in   \cite{Hohm:2010xe}. 

Next, we inspect 
(\ref{totant}), which 
determines the totally antisymmetric part of the connection. 
Specializing to the frame basis we get 
\be
\label{totantFrame}
 \sum_{A, B, C}^{cyc}  \vev{\nabla_{E_{A}}E_{B}, E_{C}} \ = \ 
 {1\over 3}   \sum_{A, B, C}^{cyc}  
 \Bigl( \vev{[E_{A}, E_{B}]_D, E_{C} } \ +  \ E_A\, \vev{ E_B, E_C} \, \Bigr)\;. 
\ee
Using (\ref{spincomp}) on the left-hand side and (\ref{Dbracket}) on the right-hand side this reads
 \be
  -\omega_{ABC}-\omega_{BCA}-\omega_{CAB} \ = \    {1\over 3}   \sum_{A, B, C}^{cyc} \Bigl(\Omega_{ABC}+
  \frac{1}{2}E_{C}{\cal G}_{AB}+E_{A}{\cal G}_{BC}\Bigr) \;.
 \ee 
Next we can use the metricity (\ref{nablaG}) in order to rewrite all derivatives of ${\cal G}$ in 
terms of $\omega$. Bringing then all $\omega$ terms to the left-hand side it is a straightforward computation 
to show that this is equivalent to 
 \be\label{totantisymm}
  \omega_{[ABC]} \ = \ -\frac{1}{3}\Omega_{[ABC]}\;, 
 \ee
in agreement with eq.~(2.32) in   \cite{Hohm:2010xe}. 

Summarizing, the constraints (1) and (2) in sec.~\ref{connectionimpl} determine the following 
connection components $\omega_{ABC}$:   
the off-diagonal projections,  
for which the first index and the second two indices belong to opposite subspaces $C_{\pm}$, 
are determined by (\ref{offconnec});  for the diagonal projections 
the part symmetric in the last two indices is determined by (\ref{nablaG}), while the totally antisymmetric 
part is determined by (\ref{totantisymm}). 
This leaves the `Hook Young tableaux' representation ${\tiny \yng(2,1)}$
undetermined. The trace part, however, is determined by constraint (3) in sec.~\ref{connectionimpl} in terms of 
the dilaton (see, e.g., 
eq.~(2.37) in \cite{Hohm:2010xe}). Thus, the undetermined part, which we denote by $\tilde{\omega}$, 
takes values in the traceless Hook representation:
\be\label{tildeomega} 
 \tilde{\omega}_{abc} \;\, : \qquad \widetilde{{\small \yng(2,1)}}\;\; , 
\ee 
and completely analogously for the second $GL(D)$. 
This will be instrumental for the analysis in sec.~\ref{physcontent} below.

\subsection{Riemann tensor}
Let us now evaluate the generalized Riemann tensor with respect 
to the frame basis and verify that it is equivalent to the Riemann tensor 
in Siegel's formalism. 
We thus want to compute  
 \be
  {\cal R}_{ABCD} \ \equiv \ {\cal R}(E_{A},E_{B},E_{C},E_{D})\;. 
 \ee
It is again convenient to introduce some notation. We write for the frame components of the 
(non-tensorial) Riemann-like tensor given in 
(\ref{test}) 
 \be\label{straightR}
 \begin{split} 
  R_{ABCD}& \ \equiv \ -R(E_{A},E_B,E_C,E_D) \ = \ -\vev{ (\nabla_{E_A}\nabla_{E_{B}}-\nabla_{E_B}\nabla_{E_{A}}-
  \nabla_{[E_A,E_B]_{C}})E_C,E_D}\\[0.5ex]
  \ &= \ \vev{ E_{A}\omega_{BC}{}^{F}\,E_{F}+\omega_{BC}{}^{F}\nabla_{E_{A}}E_{F}-(A\, \leftrightarrow \, B)+\Omega_{AB}{}^{F}
  \nabla_{E_F}E_C,E_D}\\[0.5ex]
  \ &= \ \left(E_{A}\omega_{BC}{}^{F}-E_{B}\omega_{BC}{}^{F}-\omega_{BC}{}^{E} \omega_{AE}{}^{F} 
  +\omega_{AC}{}^{E} \omega_{BE}{}^{F}-\Omega_{AB}{}^{E} \omega_{EC}{}^{F}\right){\cal G}_{FD} \;, 
 \end{split} 
 \ee  
using in the second line  (\ref{anholonomy}). The object $R_{ABCD}$ so defined
agrees  with  the object with the same name in~\cite{Hohm:2010xe}. 
The combination in the generalized Riemann tensor (\ref{curvaturehere_it_is}) contains 
the D- rather than the C-bracket, and so we  have to compute the difference. 
With (\ref{Dbracket}) we have 
  \be
  \begin{split}
  -\nabla_{[E_{A},E_{B}]_{D}}E_{C} \ &= \ 
   -\nabla_{[E_{A},E_{B}]_{C}}E_{C}
   -\frac{1}{2}E^{D}{\cal G}_{AB}\,\nabla_{E_{D}}E_{C} \\
    \ &= \ 
    -\nabla_{[E_{A},E_{B}]_{C}}E_{C}+\frac{1}{2}E^{D}{\cal G}_{AB}\,\omega_{DC}{}^{E} E_{E}\;. 
  \end{split}
 \ee 
We then conclude with (\ref{straightR})  
 \be
  \vev{ \big(\nabla_{E_{A}}\nabla_{E_{B}}-\nabla_{E_{B}}\nabla_{E_{A}} -\nabla_{[E_{A},E_{B}]_{D}} \big)E_{C},E_{D}}
   \ = \ -R_{ABCD}+\frac{1}{2}E^{E}{\cal G}_{AB}\,\omega_{ECD}\;, 
 \ee
where as usual we raise and lower frame indices with ${\cal G}_{AB}$.   
The final term in the last line in  (\ref{curvaturehere_it_is}) simply reads
 \be
  \vev{E_{B},\nabla_{E_F}E_{A}}\, \vev{E_{D},\nabla_{E^F}E_{C}} \ = \ \omega_{FAB}\,\omega^{F}{}_{CD}\;.
 \ee
The full Riemann tensor is then finally given by 
 \be
  \begin{split}
   {\cal R}(E_{A},E_{B},E_{C},E_{D}) \ = \ &-R_{ABCD}-R_{CDAB} +\omega_{EAB}\omega^{E}{}_{CD}     \\
   &+\frac{1}{2}E^{E}{\cal G}_{AB}\,\omega_{ECD}+\frac{1}{2}E^{E}{\cal G}_{CD}\,\omega_{EAB}\;.
  \end{split}
 \ee 
In order to compare with Siegel we rewrite the derivatives of ${\cal G}$ as covariant derivatives as in (\ref{nablaG}), 
which gives after a short computation 
  \be\label{frameRiemann}
  \begin{split}
   {\cal R}(E_{A},E_{B},E_{C},E_{D}) \ = \ -\Big(&R_{ABCD}+R_{CDAB}+\frac{1}{2}\omega_{ECD}\,\omega^{E}{}_{BA}
   +\frac{1}{2}\omega_{EAB}\,\omega^{E}{}_{DC}\\
   &-\frac{1}{2}\omega_{ECD}\,\nabla^{E}{\cal G}_{AB}-\frac{1}{2}\omega_{EAB}\,\nabla^{E}{\cal G}_{CD}\Big)\;.
  \end{split}
 \ee    
Comparison with eq.~(2.50) of \cite{Hohm:2010xe} 
then shows 
 \be
  {\cal R}(E_{A},E_{B},E_{C},E_{D}) \ = \ -2\,{\cal R}_{ABCD}^{\rm S}\;, 
 \ee
with ${\cal R}^{\rm S}$ denoting the Riemann tensor in the frame formalism,  
which is what we wanted to show.    
We may also impose the metricity condition for ${\cal G}$, in 
which case the second line vanishes. The resulting Riemann tensor then agrees with 
eq.~(2.48) of \cite{Hohm:2010xe}. 

Since we have now shown that the invariantly defined generalized Riemann tensor 
(\ref{curvaturehere_it_is}) reduces for a frame basis to the Riemann tensor of Siegel's 
frame formalism we can immediately derive some conclusions from our previous results. 
First, the identity (\ref{vmlck}) shows that for certain index projections the 
generalized Riemann tensor reduces to the naive one (\ref{straightR}), 
 \be
  {\cal R}_{\bar{a}b\, CD} \ = \ R_{\bar{a}b\, CD}\;.
 \ee
Second, the identity (\ref{simpcons}) implies for the frame basis 
 \be\label{STeP567}
  {\cal R}_{\bar{a}b\bar{c}d} \ \equiv \ 0\;.
 \ee
Finally, the algebraic Bianchi identity (\ref{INVAlgBianchi}) reads for 
vanishing torsion
 \be\label{STeP823}
  {\cal R}_{[ABCD]} \ = \ 0\;  \quad \Rightarrow \quad  {\cal R}_{[ABC]D} \ = \ 0\;,
 \ee
where the last implication follows as in sec.~\ref{algebraicSec1}, which in combination 
with (\ref{STeP567}) implies 
 \be\label{STeP568}
  {\cal R}_{\bar{a}\bar{b}cd} \ \equiv \ 0\;.
 \ee
It may be very tedious to verify identities like (\ref{STeP823}) and (\ref{STeP568}) using the component expression (\ref{frameRiemann}), 
but here, employing a proper geometric framework, we almost get them for free.

\subsection{Physical content of the Riemann tensor}\label{physcontent}
We now analyze to what extent the generalized Riemann tensor encodes the usual curvature invariants 
of Riemannian geometry. We will show that it contains the Ricci tensor and Ricci scalar, but that due to 
the presence of undetermined connections it does not contain the full uncontracted Riemann tensor. 

We begin by recalling that due to (\ref{STeP567}) and (\ref{STeP568}) the non-vanishing independent components of the 
Riemann tensor are 
 \be\label{RiemannList}
  {\cal R}_{abcd}\;, \quad {\cal R}_{abc\bar{d}}\;, \quad {\cal R}_{\bar{a}\bar{b}\bar{c}\bar{d}}\;, \quad {\cal R}_{\bar{a}\bar{b}\bar{c}d}\;.
 \ee 
Let us first consider ${\cal R}_{abc\bar{d}}$. Its unbarred indices  are antisymmetric in $a, b$ and therefore
belong to the $GL(D)$ representation    
   \bea
  ab,c\; :\qquad {\small \yng(1,1)}\;\otimes\; {\small \yng(1)}\hspace{0.2em}
  \ = \  {\small \yng(1,1,1)}\; \oplus\; \widetilde{{\small \yng(2,1)}} \;\oplus \;
  {\small \yng(1)} \;,  
 \eea
where the Young tableaux refer to the left $GL(D)$. The tilde in the Hook diagram 
indicates the traceless part, and the box diagram represents the trace part.   We have the algebraic Bianchi identity 
${\cal R}_{[abc]\bar{d}}\equiv 0$ and so the totally antisymmetric part is actually absent. 
In total, the full tensor ${\cal R}_{abc\bar{d}}$ belongs to the representation 
 \be\label{fullRiemprod}
{\cal R}_{abc\bar{d}}   
~\leftrightarrow  ~ 
  (ab,c)\odot  \bar{d}\; :\qquad
   \left(\;\widetilde{{\small \yng(2,1)}} \;\oplus \; {\small \yng(1)}\;\right)\;\odot\; \ \yng(1)\;. 
 \ee  
Here 
the rightmost  
box diagram refers to the right $GL(D)$, corresponding to the index $\bar{d}$, and 
we have indicated the product by $\odot$ in order to stress that the second factor belongs to a \textit{different} $GL(D)$.   
Now, the undetermined connection $\tilde{\omega}$ in (\ref{tildeomega}) makes a contribution
to the tensor ${\cal R}_{abc\bar{d}}$.  
Moreover, this  undetermined connection lives precisely in the traceless Hook diagram, 
and so its contribution to ${\cal R}_{abc\bar{d}}$
lives 
in the representation\footnote{We note that both derivative operators $E_{a}$ and $E_{\bar{a}}$ are 
non-zero even when we solve the strong constraint by setting $\tilde{\partial}^i=0$. Thus, 
all derivatives contribute a factor of $D$ additional components.}  
 \be
  {\cal R}_{abc\bar{d}}(\tilde{\omega}) \ \sim \ E_{\bar{d}}\,\tilde{\omega}_{abc}+\cdots\;\; : \qquad
  \widetilde{{\small \yng(2,1)}} \,\;\odot \; \ \yng(1) \;, 
 \ee 
where the traceless Hook diagram represents the undetermined connection and the 
second box refers again to the right $GL(D)$. In the above, 
the dots represent additional terms that may  even be free of undetermined connections. 
But, even if such terms were relevant, they are not accessible since they are affected
by the fully undetermined structure $E_{\bar{d}}\,\tilde{\omega}_{abc}$ in the representation
indicated by the above tableaux.   That full representation is therefore unavailable and, 
subtracting it  from  (\ref{fullRiemprod}),  
we conclude that the `physical' representations encoded in ${\cal R}_{abc\bar{d}}$, i.e., 
those independent of the undetermined connections, are given by ${\tiny \yng(1)}\odot {\tiny \yng(1)}$. 
This is precisely the representation content of the generalized Ricci tensor with $D^2$ components, 
 \be
  {\cal R}_{a\bar{b}} \;\; : \qquad {\small \yng(1)}\;\odot \; {\small \yng(1)}\;\, .
 \ee 
Thus we have shown that the physical content of ${\cal R}_{abc\bar{d}}$ is given 
by its trace part, the generalized Ricci tensor. 

Next we turn to the first structure in (\ref{RiemannList}), the tensor ${\cal R}_{abcd}$. It is antisymmetric in 
$a,b$ and $c,d$ and so lives in the symmetric  tensor product (compare (\ref{irreducible}))
  \bea\label{401rep}
 \left(\hspace{0.2em}{\small \yng(1,1)}\otimes {\small \yng(1,1)}\hspace{0.2em}\right)_{\rm sym}
  \ = \   {\small \yng(1,1,1,1)}\oplus 
  \widetilde{\small \yng(2,2)}\;\oplus\; \widetilde{\small \yng(2)}\;\oplus\; {\bf 1} \;,  
 \eea
where we decomposed into traceless representations, and ${\bf 1}$ denotes the singlet representation 
corresponding to the double trace. As the totally antisymmtric part is again absent due to the algebraic Bianchi identity, 
${\cal R}_{abcd}$ lives in the following representation
  \bea\label{40rep}
 {\cal R}_{abcd} \;\; : \qquad  {\small \yng(2,2)} 
  \ = \   
  \widetilde{\small \yng(2,2)}\;\oplus\; \widetilde{\small \yng(2)}\;\oplus\; {\bf 1} \;.   
 \eea
  Let us now compare with the contribution of the undetermined 
connection $\tilde{\omega}$ to ${\cal R}_{abcd}$, which reads
 \be\label{40reptilde}
   {\cal R}_{abcd}(\tilde{\omega}) \ \sim \ E_{a}\tilde{\omega}_{bcd}+\cdots \;\; : \qquad
   \left(\hspace{0.2em} \yng(1) \;\otimes\;  \widetilde{{\small \yng(2,1)}}\hspace{0.2em}\right)_{(2,2)} 
   \ = \ \widetilde{\small \yng(2,2)}\;\oplus\; \widetilde{\small \yng(2)}\;.
 \ee  
Here the subscript $(2,2)$ indicates the projection onto the representations contained in the $(2,2)$
window Young tableaux (\ref{40rep}). There is no singlet (double trace) contribution 
since  $\tilde{\omega}$ is traceless. Thus, comparing (\ref{40reptilde}) with (\ref{40rep}), we conclude that the physical 
representation encoded in this projection of the Riemann tensor (i.e., that independent of the undetermined connection) 
is precisely given by the singlet, 
 \be
   {\cal R} \;\; : \qquad {\bf 1}\;\, , 
 \ee 
which corresponds to the generalized scalar curvature.  

The analogous arguments apply to the 
final two projections in (\ref{RiemannList}), with the role of the two $GL(D)$ groups interchanged
and with respect to the undetermined connection $\tilde{\omega}_{\bar{a}\bar{b}\bar{c}}$. 
Moreover, the Ricci tensor and scalar curvature obtained from these projections are 
equivalent to those discussed above, as can be verified in an explicit basis, see e.g.~sec.~3.2 and 3.3 in  \cite{Hohm:2011si}. 
Thus, in total, the physical content of the generalized Riemann tensor is encoded 
by the generalized Ricci tensor and scalar curvature.

\section{Differential identities and the Riemann tensor}
In this section we report on some results that originated from attempts 
to derive differential Bianchi identities for the Riemann tensor beyond those 
following from the gauge invariances of an action. Although ultimately  
unsuccessful, we obtain on the way some interesting equations that may turn 
out to be useful, specifically for the gauge variation of the connection 
symbols and for a particular triple-commutator of covariant derivatives.  
Here we find it convenient to leave the invariant language and write everything in 
terms of a basis.

\subsection{Covariant gauge variation of connections}
One main difference between the geometry of double field theory and 
ordinary Riemannian geometry is that the commutator of covariant 
derivatives generally does not take a nice form in terms of the generalized 
Riemann tensor. We can write  
 \be\label{commutator}
  \big[\nabla_M,\nabla_N\big]V_{K} \ = \ -R_{MNK}{}^{L} V_{L}-T_{MN}{}^{L}\nabla_{L}V_{K}\;,
 \ee
but here we obtain the `naive' Riemann tensor (\ref{test}) and the naive torsion tensor, $T_{MN}{}^{K}=2\Gamma_{[MN]}{}^{K}$, 
which do not have tensor character (although the sum on the right-hand side of (\ref{commutator}) 
of course does \cite{Hohm:2011si}).    
Nevertheless, in the following we use this relation to rewrite the gauge variation 
of the connection symbols $\Gamma_{MNK}$ in a manifestly covariant form involving 
the proper generalized Riemann tensor. 

The gauge variation of $\Gamma$ is given by \cite{Hohm:2010xe,Hohm:2011si}
 \be\label{Gammvar}
  \delta_{\xi}\Gamma_{MNK} \ = \  \partial_M (\partial_N\xi_K-\partial_K\xi_N) +\widehat{\cal L}_{\xi}\Gamma_{MNK}\;. 
 \ee  
We replace now each partial derivative by a covariant derivative, adding and subtracting the corresponding 
connection terms. As  (\ref{Gammvar}) contains second derivatives this yields terms with 
first derivatives on connections, which we can rewrite in terms of the Riemann-like tensor $R$. 
After a straightforward computation one obtains 
\be\label{commSTEp}
\delta_\xi \Gamma_{MNK}  \ = \ 
2 \nabla_M \nabla_{[N} \xi_{K]}   + \xi^P R_{PMNK}  + T_{NK}{}^P \nabla_P \xi_M   - \Gamma_{QNK} \Gamma^Q{}_{MP} \xi^P\;.  
\ee
We can now use (\ref{commutator}) to rewrite the term on the right-hand side involving $T$, 
  \be
  T_{NK}{}^{P}\nabla_{P}\xi_{M} \ = \ -\big[\nabla_{N},\nabla_{K}\big]\xi_{M} - R_{NKMP}\xi^{P}\;. 
 \ee
Inserting this into (\ref{commSTEp}) we obtain 
 \be
\delta_\xi \Gamma_{MNK}  \ = \ 
2 \nabla_M \nabla_{[N} \xi_{K]}  -\big[\nabla_{N},\nabla_{K}\big]\xi_{M} 
+ \xi^P\big( R_{PMNK}  + R_{NKPM} + \Gamma_{QPM} \Gamma^Q{}_{NK}\big)  \;, 
\ee
using the antisymmetry in the last two indices of $R$ and $\Gamma$. 
The terms in brackets constitute  precisely the coordinate expression (\ref{calRiemcomp}) 
of the generalized Riemann tensor. 
Thus, after a minor rewriting, we have shown 
 \be\label{covGAMMA}
\delta_\xi \Gamma_{MNK}  \ = \ 
2\big( \nabla_M \nabla_{[N} \xi_{K]}   - \nabla_{[N} \nabla_{K]} \xi_M\big)   +\xi^{P}{\cal R}_{PMNK}\;. 
\ee
This is manifestly covariant, as it should be since the variation of a connection is a tensor. 
This relation is the analogue of a similar expression in Riemannian geometry, where  
 \be\label{RiemChrisVArr}
  \delta \Gamma_{mn}^{k} \ = \ 
  \nabla_m \nabla_n \xi^k  
  +   \xi^l  R_{l m}{}^k{}_n \,,    
\ee
and where $\Gamma$ denotes the usual Christoffel symbols and $R$ the usual Riemann tensor.

In the above computation we have repeatedly used the generalized torsion constraint ${\cal T}=0$. 
For completeness let us note that the variation of $\Gamma$ can also be written covariantly 
for non-vanishing torsion. After a lengthier computation, which we do not display, one obtains  
  \be\label{fullcovvar}
  \begin{split} 
\delta_\xi \Gamma_{MNK}  \ = \ 
&2\big( \nabla_M \nabla_{[N} \xi_{K]}   - \nabla_{[N} \nabla_{K]} \xi_M\big)   +\xi^{P}{\cal R}_{PMNK}\\
&
+\xi^{P}\nabla_{M}{\cal T}_{NKP}+\big(\nabla_{M}\xi^{P}-\nabla^{P}\xi_{M}\big){\cal T}_{NKP} \;, 
\end{split}  
\ee  
which reduces to (\ref{covGAMMA}) for ${\cal T}=0$.  

It is amusing to note that these relations allow us to give an alternative proof for the 
algebraic Bianchi identity (\ref{first-rw2}). We first recall that the generalized torsion tensor can be alternatively 
defined by the relation \cite{Hohm:2011si}
 \be\label{torsion}
  \big(\widehat{\cal L}_{\xi}^{\nabla}-  \widehat{\cal L}_{\xi}\big)V_{M} \ = \ {\cal T}_{MNK}\xi^{N}V^{K}\;, 
 \ee
where $\widehat{\cal L}_{\xi}^{\nabla}$ denotes the generalized Lie derivative in which each partial 
derivative has been replaced by a covariant derivative. 
This implies for any generalized vector $V$
 \be\label{LieTorsion}
  \delta_{\xi}V_{M} \ = \ \widehat{\cal L}_{\xi}V_{M} \ = \ \widehat{\cal L}_{\xi}^{\nabla}V_{M}-{\cal T}_{MN}{}^{K}V_{K}\xi^{N}\;, 
 \ee
and analogously for arbitrary generalized tensors.  
Therefore we can apply this relation to the gauge transformation of ${\cal T}_{MNK}$ itself, 
 \be\label{covtorVar}
 \begin{split}
  \delta_{\xi}{\cal T}_{MNK} \ &= \   \widehat{\cal L}_{\xi}^{\nabla}{\cal T}_{MNK}-{\cal T}_{MP}{}^{Q}{\cal T}_{QNK}\xi^{P}
  -{\cal T}_{NP}{}^{Q}{\cal T}_{MQK}\xi^{P}-{\cal T}_{KP}{}^{Q}{\cal T}_{MNQ}\xi^{P} \\
  \ &= \ \xi^{P}\nabla_{P}{\cal T}_{MNK}+3\big(\nabla_{[M}\xi^{P}-\nabla^{P}\xi_{[M}\big){\cal T}_{NK]P}
  -3{\cal T}_{[MN}{}^{Q}{\cal T}_{K]PQ}\xi^{P}\;, 
 \end{split}
 \ee
using the total antisymmetry of ${\cal T}$ in the second line.  On the other hand, we can also compute this gauge 
transformation from (\ref{fullcovvar}), using ${\cal T}_{MNK}=3\Gamma_{[MNK]}$, by simply 
projecting that equation to the totally antisymmetric part:
 \be
  \delta_{\xi}{\cal T}_{MNK} \ = \ 3\delta_{\xi}\Gamma_{[MNK]} \ = \ 3\xi^{P}{\cal R}_{P[MNK]}+3\xi^{P}\nabla_{[M}{\cal T}_{NK]P}
  +3\big(\nabla_{[M}\xi^{P}-\nabla^{P}\xi_{[M}\big){\cal T}_{NK]P}\;. 
 \ee  
Since this has to be equal to (\ref{covtorVar}) for arbitrary $\xi^{P}$ we immediately conclude
 \be
  3 {\cal R}_{P[MNK]} \ = \ \nabla_{P}{\cal T}_{MNK}-3\nabla_{[M}{\cal T}_{NK]P}-3{\cal T}_{[MN}{}^{Q}{\cal T}_{K]PQ}\;, 
 \ee
or, after a minimal rewriting, 
 \be
  3{\cal R}_{[MNK]P} \ = \ 4\nabla_{[M}{\cal T}_{NKP]}  +3{\cal T}_{[MN}{}^{Q}{\cal T}_{K]PQ}\;,
 \ee
which is the full algebraic Bianchi identity (\ref{first-rw2}).    

\medskip 

We close this section by giving the analogue of (\ref{covGAMMA}) for the spin connection coefficients. 
Their gauge variation is determined by the gauge variation of $\Gamma$ by means of the 
vielbein postulate (\ref{VielBeinPost}), 
 \be\label{VielBeinPost2}
  \nabla_{M}E_{A}{}^{N} \ = \ \partial_M E_{A}{}^{N}+\Gamma_{MK}{}^{N}E_{A}{}^{K}+\omega_{MA}{}^{B}E_{B}{}^{N} \ = \ 0\;. 
 \ee
Variation of (\ref{VielBeinPost2}) then yields 
 \be\label{covconST}
  0 \ = \ \nabla_{M}\big(\delta E_{A}{}^{N}\big)+\delta\Gamma_{MK}{}^{N} E_{A}{}^{K}+\delta\omega_{MA}{}^{B} E_{B}{}^{N}\;.
 \ee
This equation determines the $\xi$ gauge transformations of $\omega$. However, $\omega$ 
also transforms under local frame transformations corresponding to the structure group $GL(D)\times GL(D)$ 
under which $\Gamma$ is inert. It is convenient to 
determine $\delta\omega$ for a particular combination of gauge transformations with respect to $\xi$ and
a field-dependent frame transformation, setting the  $GL(D)\times GL(D)$ parameter to
 \be\label{fieldpara}
  \Lambda_{A}{}^{B} \ = \ \xi^{N}\omega_{NA}{}^{B}\;.
 \ee 
This implies for the frame components ${\cal G}_{AB}$ of the $O(D,D)$ metric 
 \be
  \begin{split}
  \delta {\cal G}_{AB} \ &= \ \xi^N \partial_N {\cal G}_{AB}+\Lambda_{A}{}^{C}{\cal G}_{CB}+\Lambda_{B}{}^{C}{\cal G}_{AC} 
  \ = \ \xi^{N}\left(\partial_N {\cal G}_{AB}+\omega_{NA}{}^{C}{\cal G}_{CB}+\omega_{NB}{}^{C}{\cal G}_{AC}\right) \\
  \ &= \ \xi^N \nabla_N {\cal G}_{AB} \ \equiv \ 0\;, 
 \end{split}
 \ee 
using the covariant constancy of ${\cal G}_{AB}$ in the last step. Thus, under this combination of gauge transformations 
${\cal G}_{AB}$ is invariant and so we can freely raise and lower frame indices inside gauge variations.  
Contracting next (\ref{covconST}) with the inverse vielbein $E_{N}{}^{C}$ we obtain after a minor rewriting and relabeling of indices
 \be\label{STEPPPP11}
  \delta \omega_{MAB} \ = \ -\nabla_{M}\left(\delta E_{A}{}^{N} E_{NB}\right)-\delta \Gamma_{MNK} E_{A}{}^{N} E_{B}{}^{K}\;,
 \ee   
where we used the relation $E_{M}{}^{A}={\cal G}^{AB}E_{B}{}^{N}\eta_{MN}$ in order to adjust the index positions, and  
the covariant constancy (\ref{VielBeinPost2}) to move the vielbein under $\nabla_{M}$. 
We next use that for the combined $\xi$ gauge transformations and field-dependent frame transformations 
with parameter (\ref{fieldpara}) we have 
 \be
  \delta E_{A}{}^{N} E_{NB} \ = \ -\left(\nabla_{A}\xi_{B}-\nabla_{B}\xi_{A}\right)\;, 
 \ee
where $\xi_{A}=E_{A}{}^{M}\xi_{M}$, as has been shown in eq.~(3.14) in \cite{Hohm:2010xe}.   
Inserting now this and (\ref{covGAMMA}) into (\ref{STEPPPP11}) we obtain 
 \be
 \begin{split}
  \delta \omega_{MAB} \ = \ &\nabla_M\left(\nabla_A\xi_B- \nabla_B\xi_A\right)\\
  &-\left(\nabla_M(\nabla_N\xi_K-\nabla_K\xi_N)-\big[\nabla_N,\nabla_K\big]\xi_M
  +\xi^P{\cal R}_{PMNK}\right)E_{A}{}^{N} E_{B}{}^{K}\;,
 \end{split} 
 \ee 
Using the covariant constancy of the vielbein, we can 
convert indices in the second line to frame indices, after which we see that  
the terms in the first line cancel. In total we get 
 \be\label{finaldelomega}
  \delta \omega_{MAB} \ = \ \big[\nabla_A,\nabla_B\big]\xi_M+{\cal R}_{ABMN}\xi^{N}\;, 
 \ee
using the symmetry properties of ${\cal R}$ in the last step. This is our final form of the gauge variation 
of the spin connection.   It is quite a remarkable relation, for the right-hand side is precisely the combination 
that would be zero if things were as in Riemannian geometry, where the commutator of covariant
derivatives yields the Riemann tensor. In contrast, here it determines the gauge 
variation of $\omega$, which is generally non-zero. There is one exception, however.
The special 
component $\omega_{Ma\bar{b}}$ vanishes and hence its variation vanishes, which implies 
 \be
  \big[\nabla_a,\nabla_{\bar{b}}\big]\,\xi_M \ = \ -{\cal R}_{a\bar{b}\,MN}\,\xi^{N}\;. 
 \ee 
It is clear from (\ref{finaldelomega}) that this is the only simple relation between the commutator of covariant derivatives 
and the generalized Riemann tensor.

\subsection{Triple commutators and the Riemann tensor}
We have seen above that (\ref{fullcovvar}) allows for a simple proof of the algebraic Bianchi identity. 
One may thus wonder whether we can also obtain a differential Bianchi identity this way. 
Indeed, in conventional Riemannian geometry the differential Bianchi identity can be 
proved along these lines, which we briefly sketch in the following. The gauge transformation of the Christoffel symbols $\Gamma_{mn}^{k}$ 
can be written as in (\ref{RiemChrisVArr}). 
Using this in the general variation of the Riemann tensor,  
 \be\label{DelR}
  \delta R_{mn}{}^{k}{}_{l} \ = \ \nabla_{m}\delta\Gamma_{nl}^{k}-
  \nabla_{n}\delta\Gamma_{ml}^{k}\;, 
 \ee
employing the standard relation $ [\nabla_{m},\nabla_{n}]V^{k}  =  R_{mn}{}^{k}{}_{l}V^{l}$,  
one finds after a brief computation 
 \be\label{delRiemann1}
  \delta R_{mn}{}^{k}{}_{l} \ = \  -2\xi^{p}\nabla_{[m}R_{n]p}{}^{k}{}_{l}
  -2\nabla_{[m}\xi^{p}\,R_{n]p}{}^{k}{}_{l}-\nabla_{p}\xi^{k}R_{mn}{}^{p}{}_{l}
  + \nabla_{l}\xi^{p} R_{mn}{}^{k}{}_{p}\;. 
 \ee     
On the other hand, this gauge transformation can also be written as the standard 
Lie derivative, but with all partial derivatives replaced by covariant derivatives 
(as we are allowed to do for the Levi-Civita connection), 
    \be
  \delta R_{mn}{}^{k}{}_{l} \ = \ \xi^{p}\nabla_{p} R_{mn}{}^{k}{}_{l}
  -2\nabla_{[m}\xi^{p}\,R_{n]p}{}^{k}{}_{l}-\nabla_{p}\xi^{k}\,R_{mn}{}^{p}{}_{l}
  +\nabla_{l}\xi^{p}\,R_{mn}{}^{k}{}_{p}\;.
 \ee
As this has to agree with (\ref{delRiemann1}) for arbitrary $\xi$ we conclude 
 \be
   3\nabla_{[p}R_{mn]}{}^{k}{}_{l} \ = \ \nabla_{p} R_{mn}{}^{k}{}_{l} + 2\nabla_{[m}R_{n]p}{}^{k}{}_{l}  \ \equiv  \ 0\;, 
 \ee
which is the differential Bianchi identity we wanted to prove.

A crucial step in the above proof, from (\ref{DelR}) to (\ref{delRiemann1}), 
was to rewrite the commutator of covariant derivatives with the Riemann tensor. 
As discussed above, in the generalized case we do not have such an identity, so 
it appears doubtful whether the above strategy can be employed. However, what is really needed
are triple-commutators of covariant derivatives, and it turns out that there 
is such an identity in terms of the generalized Riemann tensor. Unfortunately, following the above 
steps it leads to $0=0$, thus confirming the suspicion expressed in \cite{Hohm:2011si} 
that there is no analogue of an uncontracted differential Bianchi identity. 
Rather, we now turn the logic around and use this observation to give a simple 
proof for this triple-commutator relation, which may be
useful for other applications. 

We start with the general variation of the generalized Riemann tensor, which can be written similar 
to the standard case \cite{Hohm:2011si}, 
 \be\label{GenVarRiem}
  \delta{\cal R}_{MNKL} \ = \ 2\nabla_{[M}\delta\Gamma_{N]KL}+2\nabla_{[K}\delta\Gamma_{L]MN}\;. 
 \ee
We can now specialize to the generalized diffeomorphism transformation in the form (\ref{covGAMMA}),  
which yields after a brief computation 
  \be\label{DElR}
 \begin{split}
  \delta   &{\cal R}_{MNKL}  \ = \ 2\Big(\big[ \big[\nabla_{M},\nabla_{N}\big],\nabla_{[K}\big]\xi_{L]}+\big[\big[\nabla_{K},\nabla_{L}\big],\nabla_{[M}\big]\xi_{N]}\\
  &\qquad \,
  +\nabla_{[M}\xi^{P}\,{\cal R}_{|P|N]KL}+\nabla_{[K}\xi^{P}\,{\cal R}_{|P|L]MN}+\xi^{P}\big(\nabla_{[M}{\cal R}_{|P|N]KL}
  +\nabla_{[K}{\cal R}_{|P|L]MN}\big)\Big)\;. 
 \end{split}
 \ee
As above we know that this must be equal to the generalized Lie derivative, and for vanishing torsion 
we can replace all partial derivatives by covariant derivatives, see (\ref{LieTorsion}). Thus, we have 
  \be
  \begin{split}
  \delta_{\xi}{\cal R}_{MNKL} \ = \ &\xi^{P}\nabla_{P}{\cal R}_{MNKL}\\
  &+2\big(\nabla_{[M}\xi^{P}-\nabla^{P}\xi_{[M}\big){\cal R}_{|P|N]KL}
  +2\big(\nabla_{[K}\xi^{P}-\nabla^{P}\xi_{[K}\big){\cal R}_{|P|L]MN} \;. 
 \end{split}
 \ee 
 Comparing this with (\ref{DElR}) we infer the following triple-commutator relation valid for an arbitrary vector
 $\xi$:   
  \be\begin{split}
  \label{triplecomm}
     \Big[ \big[\nabla_{M},\nabla_{N}\big],\nabla_{[K}\Big]\xi_{L]}+\Big[\big[\nabla_{K},\nabla_{L}\big],\nabla_{[M}\Big]\xi_{N]} 
     \ = & \ \   
      {\cal R}_{MNP[K}\nabla^{P}\xi_{L]}+{\cal R}_{KLP[M}\nabla^{P}\xi_{N]} 
  \\
  & \hskip-80pt  +\xi^{P}\Big(\nabla_{[M}{\cal R}_{N]PKL}+\nabla_{[K}{\cal R}_{L]PMN}+\frac{1}{2}\nabla_{P}{\cal R}_{MNKL}\Big)\;. 
   \end{split}\ee

\subsection{Differential Bianchi identities from higher-derivative actions}
Although we have argued in \cite{Hohm:2011si} and above that there is no analogue of the 
differential Bianchi identity $\nabla_{[m}R_{nk]pq}=0$ in double field theory, there is of course the Bianchi identity 
following from the gauge invariance of the double field theory action  (\ref{DFTaction}), 
which in turn is a generalization of the usual Bianchi identity $\nabla^m G_{mn}=0$ for the 
Einstein tensor in general relativity.  Similarly, one can derive further differential Bianchi identities 
by using the gauge invariance of higher-derivative actions such as 
 \be\label{RiemSQ}
  S \ = \ \int dxd\tilde{x}\,e^{-2d}\, {\cal R}^{MNKL}\, {\cal R}_{MNKL}\;.
 \ee
This action as such is not of direct physical interest, for it depends on undetermined 
connections and thus involves more than the physical fields. Moreover, as we showed 
in the previous section, it does not contain the (square of the) 
Riemann tensor and is therefore insufficient for 
$\alpha'$ corrections. All we need, however, is the gauge invariance of (\ref{RiemSQ}), which is manifest. 

Let us now derive the differential Bianchi identity.  
This requires (\ref{GenVarRiem}) and (\ref{covGAMMA}) for the variation of the 
Riemann tensor and \cite{Hohm:2011si}
 \be
  \delta_{\xi} e^{-2d} \ = \ e^{-2d}\nabla_P\xi^P
 \ee
for the gauge transformation of the dilaton.   We compute 
 \be
  \begin{split}
   0 \ &= \ \delta_{\xi}S \ = \ \int dxd\tilde{x}\,e^{-2d}\, \Big[\nabla_P\xi^P\,{\cal R}^{MNKL}\, {\cal R}_{MNKL} 
   +8 {\cal R}^{MNKL}\nabla_M \delta\Gamma_{NKL}\Big] \\
   \ &= \ \int dxd\tilde{x}\,e^{-2d}\, \Big[-2\xi^P{\cal R}^{MNKL}\nabla_P {\cal R}_{MNKL}\\
   &\qquad\qquad\qquad\quad\;\;   -8 \nabla_M{\cal R}^{MNKL}\big( 2\nabla_N\nabla_K\xi_L -2\nabla_K\nabla_L \xi_N +\xi^P {\cal R}_{PNKL}  \big)\Big] \\
   \ &= \ -2\int dxd\tilde{x}\,e^{-2d}\, \Big[\xi^P{\cal R}^{MNKL}\nabla_P {\cal R}_{MNKL}\\
   &\qquad\quad  +\xi^P \big(8\nabla^K\nabla^N\nabla^M {\cal R}_{MNKP}-8\nabla^L\nabla^K\nabla^M{\cal R}_{MPKL}
   + 4\nabla_M{\cal R}^{MNKL}{\cal R}_{PNKL}\big)\Big] \\
   \ &= \ -2\int dxd\tilde{x}\,e^{-2d}\ \xi^P \Big[{\cal R}^{MNKL}\nabla_P {\cal R}_{MNKL} +8\nabla^K\nabla^N\nabla^M\big({\cal R}_{MNKP}-{\cal R}_{MPNK}\big)\\
   &\qquad\qquad\qquad\qquad\qquad \;\,
   +4{\cal R}_{PNKL}\nabla_{M}{\cal R}^{MNKL}\Big]\;.
  \end{split}
 \ee   
Here we did several integrations by part and index relabelings.  
As this integral vanishes for arbitrary $\xi^P$ we conclude 
 \be
  {\cal R}_{PNKL}\nabla_{M}{\cal R}^{MNKL}+2\nabla^K\nabla^N\nabla^M\big({\cal R}_{MNKP}-{\cal R}_{MPNK}\big)
  +\frac{1}{4}{\cal R}^{MNKL}\nabla_P {\cal R}_{MNKL} \ = \ 0\;.
 \ee
We close this section by noting that along the same lines one could easily derive more differential 
Bianchi identities, by using different higher-derivative actions, e.g., involving the generalized 
Riemann tensor with certain index projections rather than the full unprojected one in (\ref{RiemSQ}).

\section{Concluding remarks} 

Even though a 
geometric framework for double field theory 
is by now quite well-understood 
in `index-based' physics terminology,  see e.g.~\cite{Hohm:2010xe,Hohm:2011si}, 
a more invariant treatment, analogous to 
the coordinate-free formulation 
of 
differential geometry developed in pure mathematics in the mid 20th century, was lacking.  
In this paper we believe to have taken a 
first step towards a  similar  
formulation of the geometry of 
double field theory. Among other things, this formulation makes 
manifest the equivalence of the 
previously developed frame-like \cite{Siegel:1993th,Hohm:2010xe} and metric-like formalisms \cite{Hohm:2011si,Jeon:2010rw,Jeon:2011cn}.  
The 
geometric structures emerging in double field theory are a 
compelling generalization of those in 
Riemannian geometry. Nevertheless,  
 there are also puzzling features which seem to suggest that the 
present framework may eventually become just part of a more general structure. 

Most importantly, in contrast to Riemannian geometry, the connection is not uniquely determined in terms 
of the physical fields.   
This is essentially because the constraint of zero generalized
torsion in double geometry
is 
weaker than the similar constraint in Riemannian geometry.  More precisely,
in Riemannian geometry the constraint of zero torsion  and the
constraint that $\nabla$ is  compatible with
the metric $\langle \cdot , \cdot \rangle$ determine completely the connection.  In the doubled geometry the analogous
conditions use the generalized torsion and the compatibility of $\nabla$ with the 
metric $\langle \cdot , \cdot \rangle$ described by $\eta$. They come up short to fix the connection 
because, when $\nabla$ is compatible with the metric, the generalized
torsion is totally antisymmetric
in its three indices and thus has far fewer components than expected.  Not even the imposition of further conditions 
-- the compatibility of $\nabla$ with the generalized metric and a trace condition on the connection -- can 
make up for the shortcoming.   With undetermined connections, we have a generalized
Riemann tensor with undetermined components.

Another puzzling feature of the generalized geometric formalism, 
perhaps related to the presence
of undetermined connection components, is the apparent absence of differential Bianchi identities beyond those 
following from the gauge invariance of actions. Specifically, it seems quite clear that there is no analogue  
of the uncontracted differential Bianchi identity for the 
Riemann tensor, but we also have not been able to 
find a once-contracted  
Bianchi identity that would generalize the familiar relation $\nabla^m R_{mnkl}=2\nabla_{[k}R_{l]n}$. 
One may wonder whether this somehow hints at the 
need to introduce some larger structures in which the significance of these observations will become clear.

While it has been known that the generalized Riemann tensor does encode the  Ricci tensor and Ricci scalar,
we have shown 
(sec.~\ref{physcontent})  that the presence of undetermined connections
implies that  the generalized Riemann tensor does not have 
enough  physical components to describe the Riemann tensor.  In fact, 
it has only enough of Riemann to determine the Ricci and scalar curvatures. 
An immediate consequence of these results is that the present framework is insufficient 
to describe 
$\alpha'$-corrections to the effective action, for the latter are known  
to include higher powers of the full Riemann tensor. 
This implies that even if there were 
a procedure to remove the undetermined connections 
from curvature-squared terms, as we speculated in \cite{Hohm:2011si},  it would not describe the 
complete couplings required by string theory to higher order in $\alpha'$.

We close be arguing that, despite the apparent complications,  there are 
very good reasons to believe that it must be possible to 
encode $\alpha'$ corrections in double field theory. For instance, in closed string field theory~\cite{Zwiebach:1992ie}, 
which uses doubled coordinates~\cite{Kugo:1992md} and inspired the recent progress in double field theory, 
T-duality is known to be present to all orders in $\alpha'$. Moreover, as shown by Sen~\cite{Sen:1991zi}, such
symmetry implies the continuous $O(D,D)$ symmetry of the effective low-energy action
to all orders in $\alpha'$, which has been verified explicitly by 
Meissner to first order in $\alpha'$ 
in reductions to one dimension \cite{Meissner:1996sa}. 
String field theory also suggests what may 
be needed in order to overcome the obstacles found here. In fact, the gauge transformations and the 
bracket governing the gauge algebra receive $\alpha'$ corrections in string field theory. Therefore 
it would not be surprising if we are forced to go beyond the geometric framework developed so far, 
e.g., generalizing the Courant and Dorfman brackets.  If a further generalization encompassing 
$\alpha'$ corrections does exist, it is reasonable to expect that the geometric structures 
discussed here will become a natural part of that larger framework. 

Apart from the problem of $\alpha'$ corrections there is a wealth of questions related
 to the pure two-derivative theory. 
Most importantly, we are still lacking a 
proper `intrinsic' understanding of the doubled manifold and the generalized vectors.   
Our recent paper \cite{Hohm:2012gk} interprets finite gauge transformations as generalized coordinate 
transformations and so suggests a generalized notion of manifold, but a
precise mathematical 
formulation and nontrivial examples are 
clearly desirable. 
In particular, such progress is needed 
to understand 
global aspects of generalized manifolds, e.g., in order to investigate 
to what extent solutions can be patched together to a globally non-trivial space. 
We hope that these questions  
will be answered in the near future.

\section*{Acknowledgments}
We would like to thank Marco Gualtieri, 
Martin Rocek, Ashoke Sen, Warren Siegel, and Mikhail Vasiliev for
helpful discussions.  

This work is supported by the U.S. Department of Energy (DoE) under the cooperative
research agreement DE-FG02-05ER41360, the
DFG Transregional Collaborative Research Centre TRR 33
and the DFG cluster of excellence "Origin and Structure of the Universe".


\begin{thebibliography}{99}
\bibitem{Hull:2009mi}
  C.~Hull, B.~Zwiebach,
  ``Double Field Theory,''
  JHEP {\bf 0909}, 099 (2009).
  [arXiv:0904.4664 [hep-th]],
\bibitem{Hull:2009zb}
  C.~Hull, B.~Zwiebach,
  ``The Gauge algebra of double field theory and Courant brackets,''
  JHEP {\bf 0909}, 090 (2009).
  [arXiv:0908.1792 [hep-th]].

\bibitem{Hohm:2010jy}
  O.~Hohm, C.~Hull and B.~Zwiebach,
  ``Background independent action for double field theory,''
  JHEP {\bf 1007} (2010) 016
  [arXiv:1003.5027 [hep-th]].

\bibitem{Hohm:2010pp}
  O.~Hohm, C.~Hull and B.~Zwiebach,
  ``Generalized metric formulation of double field theory,''
  JHEP {\bf 1008} (2010) 008
  [arXiv:1006.4823 [hep-th]].

\bibitem{Siegel:1993th}
  W.~Siegel,
  ``Superspace duality in low-energy superstrings,''
  Phys.\ Rev.\  D {\bf 48}, 2826 (1993)
  [arXiv:hep-th/9305073],
  ``Two vierbein formalism for string inspired axionic gravity,''
  Phys.\ Rev.\  D {\bf 47}, 5453 (1993)
  [arXiv:hep-th/9302036].

  \bibitem{Tseytlin:1990nb}
A.~A.~Tseytlin,
``Duality Symmetric Formulation Of String World Sheet Dynamics,''
Phys.\ Lett.\ B {\bf 242}, 163 (1990);
``Duality Symmetric Closed String Theory And Interacting Chiral Scalars,''
Nucl.\ Phys.\ B {\bf 350}, 395 (1991).

\bibitem{Duff:1989tf}
  M.~J.~Duff,
  ``Duality Rotations In String Theory,''
  Nucl.\ Phys.\ B {\bf 335}, 610 (1990),
  M.~J.~Duff and J.~X.~Lu,
  ``Duality Rotations In Membrane Theory,''
  Nucl.\ Phys.\ B {\bf 347}, 394 (1990).

\bibitem{Hohm:2010xe}
  O.~Hohm, S.~K.~Kwak,
  ``Frame-like Geometry of Double Field Theory,''
  J.\ Phys.\ A {\bf A44}, 085404 (2011).
  [arXiv:1011.4101 [hep-th]],

\bibitem{Kwak:2010ew}
  S.~K.~Kwak,
  ``Invariances and Equations of Motion in Double Field Theory,''
  JHEP {\bf 1010} (2010) 047
  [arXiv:1008.2746 [hep-th]].
\bibitem{Hohm:2011gs}
  O.~Hohm,
  ``T-duality versus Gauge Symmetry,''
  arXiv:1101.3484 [hep-th], 
   B.~Zwiebach,
  ``Double Field Theory, T-Duality, and Courant Brackets,''
  [arXiv:1109.1782 [hep-th]].

\bibitem{Hohm:2011dz}
  O.~Hohm,
  ``On factorizations in perturbative quantum gravity,''
  JHEP {\bf 1104}, 103 (2011).
  [arXiv:1103.0032 [hep-th]].

\bibitem{Hohm:2011ex}
  O.~Hohm, S.~K.~Kwak,
  ``Double Field Theory Formulation of Heterotic Strings,''
  JHEP {\bf 1106}, 096 (2011).
  [arXiv:1103.2136 [hep-th]].

\bibitem{Hohm:2011zr}
  O.~Hohm, S.~K.~Kwak, B.~Zwiebach,
  ``Unification of Type II Strings and T-duality,''
  Phys.\ Rev.\ Lett.\  {\bf 107}, 171603 (2011),
    [arXiv:1106.5452 [hep-th]],
  ``Double Field Theory of Type II Strings,''
  JHEP {\bf 1109}, 013 (2011),
    [arXiv:1107.0008 [hep-th]].

\bibitem{Hohm:2011cp} 
  O.~Hohm and S.~K.~Kwak,
  ``Massive Type II in Double Field Theory,''
  JHEP {\bf 1111}, 086 (2011)
  [arXiv:1108.4937 [hep-th]].

\bibitem{Hohm:2011nu} 
  O.~Hohm and S.~K.~Kwak,
  ``N=1 Supersymmetric Double Field Theory,''
  JHEP {\bf 1203}, 080 (2012)
  [arXiv:1111.7293 [hep-th]].

\bibitem{Hohm:2011si}
  O.~Hohm and B.~Zwiebach,
  ``On the Riemann Tensor in Double Field Theory,''
  JHEP {\bf 1205} (2012) 126
  [arXiv:1112.5296 [hep-th]].

\bibitem{Hillmann:2009ci}
  C.~Hillmann,
  ``Generalized E(7(7)) coset dynamics and D=11 supergravity,''
  JHEP {\bf 0903}, 135 (2009).
  [arXiv:0901.1581 [hep-th]].

\bibitem{Berman:2010is}
  D.~S.~Berman, M.~J.~Perry,
  ``Generalized Geometry and M theory,''
  JHEP {\bf 1106}, 074 (2011).
  [arXiv:1008.1763 [hep-th]],
  D.~S.~Berman, H.~Godazgar, M.~J.~Perry,
  ``SO(5,5) duality in M-theory and generalized geometry,''
  Phys.\ Lett.\  {\bf B700}, 65-67 (2011).
  [arXiv:1103.5733 [hep-th]],
    D.~S.~Berman, H.~Godazgar, M.~Godazgar, M.~J.~Perry,
  ``The Local symmetries of M-theory and their formulation in generalised geometry,''
  [arXiv:1110.3930 [hep-th]].

\bibitem{Berman:2012uy} 
  D.~S.~Berman, E.~T.~Musaev, and D.~C.~Thompson,
  ``Duality Invariant M-theory: Gauged supergravities and Scherk-Schwarz reductions,''
  JHEP {\bf 1210}, 174 (2012)
  [arXiv:1208.0020].

\bibitem{Berman:2012vc} 
  D.~S.~Berman, M.~Cederwall, A.~Kleinschmidt and D.~C.~Thompson,
  ``The gauge structure of generalised diffeomorphisms,''
  arXiv:1208.5884 [hep-th].


\bibitem{Jeon:2010rw}
  I.~Jeon, K.~Lee, J.~-H.~Park,
  ``Differential geometry with a projection: Application to double field theory,''
  JHEP {\bf 1104}, 014 (2011).
  [arXiv:1011.1324 [hep-th]].

\bibitem{Jeon:2011cn}
  I.~Jeon, K.~Lee, J.~-H.~Park,
  ``Stringy differential geometry, beyond Riemann,''
  Phys.\ Rev.\  {\bf D84}, 044022 (2011), 
  [arXiv:1105.6294 [hep-th]]. 
 


\bibitem{Jeon:2011sq}
  I.~Jeon, K.~Lee and J.~-H.~Park,
  ``Supersymmetric Double Field Theory: Stringy Reformulation of Supergravity,''
  arXiv:1112.0069 [hep-th], 
  ``Ramond-Ramond Cohomology and O(D,D) T-duality,''
  JHEP {\bf 1209}, 079 (2012)
  arXiv:1206.3478 [hep-th], I.~Jeon, K.~Lee, J.~-H.~Park and Y.~Suh,
  ``Stringy Unification of Type IIA and IIB Supergravities under N=2 D=10 Supersymmetric Double Field Theory,''
  arXiv:1210.5078 [hep-th].

\bibitem{Schulz:2011ye} 
  M.~B.~Schulz,
  ``T-folds, doubled geometry, and the SU(2) WZW model,''
  JHEP {\bf 1206}, 158 (2012)
  [arXiv:1106.6291 [hep-th]].

\bibitem{Copland:2011yh}
  N.~B.~Copland,
  ``Connecting T-duality invariant theories,''
  Nucl.\ Phys.\  {\bf B854}, 575-591 (2012).
  [arXiv:1106.1888 [hep-th]],
  ``A Double Sigma Model for Double Field Theory,''
  JHEP {\bf 1204}, 044 (2012), 
  [arXiv:1111.1828 [hep-th]].

\bibitem{Thompson:2011uw}
  D.~C.~Thompson,
  ``Duality Invariance: From M-theory to Double Field Theory,''
  JHEP {\bf 1108}, 125 (2011).
  [arXiv:1106.4036 [hep-th]].

\bibitem{Albertsson:2011ux}
  C.~Albertsson, S.~-H.~Dai, P.~-W.~Kao, F.~-L.~Lin,
  ``Double Field Theory for Double D-branes,''
  JHEP {\bf 1109}, 025 (2011), 
  [arXiv:1107.0876 [hep-th]].

\bibitem{Andriot:2011uh}
  G.~Aldazabal, W.~Baron, D.~Marques, C.~Nunez,
  ``The effective action of Double Field Theory,''
  JHEP {\bf 1111}, 052 (2011).
  [arXiv:1109.0290 [hep-th]],
  D.~Geissbuhler,
  ``Double Field Theory and N=4 Gauged Supergravity,''
  [arXiv:1109.4280 [hep-th]].
  
 \bibitem{grana-marques}
   M.~Grana and D.~Marques,
  ``Gauged Double Field Theory,''
  JHEP {\bf 1204}, 020 (2012)
  [arXiv:1201.2924 [hep-th]].


\bibitem{Andriot:2012wx} 
  D.~Andriot, O.~Hohm, M.~Larfors, D.~Lust and P.~Patalong,
  ``A geometric action for non-geometric fluxes,''
  Phys.\ Rev.\ Lett.\  {\bf 108}, 261602 (2012)
  [arXiv:1202.3060 [hep-th]], 
  ``Non-Geometric Fluxes in Supergravity and Double Field Theory,''
  Fortschritte der Physik, 
Volume 60, Issue 11-12, 1150�1186, 2012, arXiv:1204.1979 [hep-th]. 

\bibitem{Hohm:2012gk} 
  O.~Hohm and B.~Zwiebach,
  ``Large Gauge Transformations in Double Field Theory,''
  JHEP {\bf 1302}, 075 (2013)
  [arXiv:1207.4198 [hep-th]].
  
    
\bibitem{Hitchin:2004ut} 
  N.~Hitchin,
  ``Generalized Calabi-Yau manifolds,''
  Quart.\ J.\ Math.\ Oxford Ser.\  {\bf 54}, 281 (2003)
  [math/0209099 [math-dg]].
  
\bibitem{Gualtieri:2003dx} 
  M.~Gualtieri,
  ``Generalized complex geometry,''
  math/0401221 [math-dg].
  
\bibitem{Gualtieri:2007bq} 
  M.~Gualtieri,
  ``Branes on Poisson varieties,''
  arXiv:0710.2719 [math.DG].
  
\bibitem{Coimbra:2011nw}
  A.~Coimbra, C.~Strickland-Constable, D.~Waldram,
  ``Supergravity as Generalised Geometry I: Type II Theories,''
    [arXiv:1107.1733 [hep-th]],
  ``$E_{d(d)} \times \mathbb{R}^+$ Generalised Geometry, Connections and M theory,''
  arXiv:1112.3989 [hep-th].

  
  
\bibitem{Vaisman:2012ke} 
  I.~Vaisman,
  ``On the geometry of double field theory,''
  J.\ Math.\ Phys.\  {\bf 53}, 033509 (2012)
  [arXiv:1203.0836 [math.DG]], 
    ``Towards a double field theory on para-Hermitian manifolds,''
  arXiv:1209.0152 [math.DG].
  
\bibitem{Zwiebach:1992ie} 
  B.~Zwiebach,
  ``Closed string field theory: Quantum action and the B-V master equation,''
  Nucl.\ Phys.\ B {\bf 390}, 33 (1993)
  [hep-th/9206084].
  
\bibitem{Kugo:1992md} 
  T.~Kugo and B.~Zwiebach,
  ``Target space duality as a symmetry of string field theory,''
  Prog.\ Theor.\ Phys.\  {\bf 87}, 801 (1992)
  [hep-th/9201040].

\bibitem{Sen:1991zi} 
  A.~Sen,
  ``O(d) x O(d) symmetry of the space of cosmological solutions in string theory, scale factor duality and two-dimensional black holes,''
  Phys.\ Lett.\ B {\bf 271}, 295 (1991).

\bibitem{Meissner:1996sa} 
  K.~A.~Meissner,
  ``Symmetries of higher order string gravity actions,''
  Phys.\ Lett.\ B {\bf 392}, 298 (1997)
  [hep-th/9610131].

\end{thebibliography}
\end{document}